f% ============================================================
\documentclass[11pt,peerreview,a4paper,draft]{IEEEtran}

\usepackage[T1]{fontenc}
\usepackage[english]{babel}
\usepackage[active]{srcltx}
\usepackage{color}
\usepackage{amsmath}
\usepackage{amssymb}
\usepackage[final]{graphicx}
\usepackage{amsbsy}
\usepackage{url}
\usepackage{enumerate}
\usepackage{enumerate}
\usepackage{euscript}
\usepackage{mathrsfs}
\usepackage{psfrag}
\usepackage{latexsym}

\renewcommand{\tilde}{\widetilde}
\newcommand{\bm}[1]{{\boldsymbol{\rm #1}}}
\newcommand{\Es}{{\mathbb{E}}}

\newcommand{\diag}{{\text{diag}}}
\newcommand{\trace}{{\text{tr}}}
\newcommand{\I}{\bm{I}}
\newcommand{\A}{\bm{A}}
\newcommand{\K}{\bm{K}}
\renewcommand{\a}{\bm{a}}
\newcommand{\B}{\bm{B}}
\newcommand{\F}{\bm{F}}
\newcommand{\bCcal}{\bm{\Psi}}
\newcommand{\Zero}{\bm{O}}

\newcommand{\Cset}{\mathbb{C}}
\newcommand{\Rset}{\mathbb{R}}
\newcommand{\Nset}{\mathbb{N}}
\newcommand{\Zset}{\mathbb{Z}}
\newcommand{\CN}{\mathcal{CN}}

\newcommand{\eqdef}{\triangleq}

\newcommand{\bpsi}{\bm{\psi}}
\renewcommand{\det}{{\mathrm{det}}}
\newcommand{\herm}{\text{H}}
\newcommand{\trasp}{\text{T}}
\def\MI{\mathsf{I}}

\newcommand{\betamax}{\beta_\text{max}}
\newcommand{\deltamin}{\delta_\text{min}}
\renewcommand{\r}{\bm{r}}
\renewcommand{\v}{\bm{v}}
\newcommand{\s}{\bm{s}}
\newcommand{\p}{\bm{p}}
\newcommand{\x}{\bm{x}}
\newcommand{\y}{\bm{y}}

\renewcommand{\d}{\bm{d}}
\renewcommand{\c}{\bm{c}}
\renewcommand{\u}{\bm{u}}

\newcommand{\C}{\bm{C}}
\newcommand{\R}{\bm{R}}
\def\Cap{\mathsf{C}}
\newcommand{\Lcp}{{L_{\text{cp}}}}
\newcommand{\Ts}{{T_{\text{s}}}}
\newcommand{\Tc}{{T_{\text{c}}}}
\newcommand{\Widft}{\mathbf{W}_{\text{IDFT}}}
\newcommand{\Wdft}{\mathbf{W}_{\text{DFT}}}
\renewcommand{\Widft}{\mathbf{W}_{\text{IDFT}}}
\newcommand{\Tcp}{\mathbf{T}_{\text{cp}}}
\newcommand{\bSigma}{\bm{\Sigma}}
\newcommand{\bXi}{\bm{\Xi}}
\newcommand{\bxi}{\bm{\xi}}
\newcommand{\bbeta}{\bm{\beta}}
\newcommand{\blambda}{\bm{\lambda}}
\newcommand{\bOmega}{\bm{\Omega}}
\def\ent{\mathsf{h}}
\def\entdis{\mathsf{H}}
\def\fish{\mathsf{J}}
\def\MSE{\Lambda_{\text{mse}}}

\newcommand{\Pmax}{\EuScript{P}_{\text{max}}^{\text{c}}}
\newcommand{\Phq}{\EuScript{P}^{\text{c}}_q}
\def\Ps{\sigma_{s}^2}
\def\Pb{\sigma_{b}^2}
\def\SNRL{\text{SNR}_{\text{L}}}
\def\SNRB{\text{SNR}_{\text{B},1}}
\def\SNRBB{\text{SNR}_{\text{B},4}}

\def\bdm#1\edm{\begin{displaymath}#1\end{displaymath}}
\def\be#1\ee{\begin{equation}#1\end{equation}}
\def\barr#1\earr{\begin{align}#1\end{align}}

\newcommand{\IeeeCOMMMAG}{{\em IEEE Commun.\ Magazine\/}}
\newcommand{\IeeeTIT}{{\em IEEE Trans.\ Inf.\ Theory\/}}
\newcommand{\IeeeTSP}{{\em IEEE Trans.\ Signal Process.\/}}
\newcommand{\IeeeTCOMM}{{\em IEEE Trans.\ Commun.\/}}

\newcommand{\IeeeJSAC}{{\em IEEE J.\ Select.\ Areas Commun.\/}}
\newcommand{\IeeeTVT}{{\em IEEE Trans.\ Veh.\ Technol.\/}}
\newcommand{\IeeeTMTT}{{\em IEEE Trans.\ Microw.\ Theory Tech.\/}}
\newcommand{\IeeeSPMAG}{{\em IEEE Signal Processing Magazine\/}}

\newcommand{\sfootnote}[1]{{\footnote{\setlength{\baselineskip}{4.0mm}#1}}}

\begin{document}

\title{
\setlength{\baselineskip}{10.0mm}
Achievable information rates of ambient \\ backscatter communications}

%\author{Donatella Darsena,~\IEEEmembership{Member,~IEEE,}
%         Giacinto Gelli, and
%        Francesco Verde,~\IEEEmembership{Senior Member,~IEEE}
%
\author{Donatella Darsena, Giacinto Gelli, and Francesco Verde
\thanks{
\setlength{\baselineskip}{4.0mm}
D.~Darsena is with the Department of Engineering,
Parthenope University, Naples I-80143, Italy (e-mail: darsena@uniparthenope.it).
G.~Gelli and F.~Verde are with the
Department of Electrical Engineering and
Information Technology, University Federico II, Naples I-80125,
Italy [e-mail: (gelli,f.verde)@unina.it].}
}

\maketitle
\vspace{-10mm}
\begin{abstract}

\setlength{\baselineskip}{5.0mm}
Ambient backscatter is an intriguing wireless communication paradigm that
allows small devices to compute and communicate by using
only the power they harvest from radio-frequency (RF) signals in the air.
Ambient backscattering devices reflect existing RF signals emitted
by legacy communications systems, such as digital TV
broadcasting, cellular or Wi-Fi ones, which would be
otherwise treated as harmful sources of interference.
This paper deals with the ultimate performance limits of ambient
backscatter systems in  broadband  fading  environments,
by considering different amounts of network state information
at the receivers. After introducing  a  detailed signal model of
the  relevant communication links,  we study the influence of
physical parameters on the capacity
of both legacy and backscatter systems.
We find that, under reasonable operative conditions,
a legacy system employing multicarrier modulation 
can turn the RF interference arising
from the backscatter process into a form of multipath
diversity that can be suitably exploited to noticeably increase 
its performance.
Moreover, we show that, even when employing 
simple single-carrier modulation techniques, the backscatter system can achieve significant data rates over relatively short distances, especially when
the intended recipient of the backscatter signal is co-located with
the legacy transmitter, i.e., they are on the same machine. 

\end{abstract}

\vspace{-8mm}
\begin{IEEEkeywords}
\vspace{-2mm}
\setlength{\baselineskip}{5.0mm}
Ambient backscatter, ergodic and outage capacity,
symbol variance and amplitude constraints, multicarrier systems,
performance bounds.
\end{IEEEkeywords}

\vspace{-5mm}

\section{Introduction}
\label{sec:intro}

\IEEEPARstart{E}{lectromagnetic} (EM) interference, also called radio-frequency
(RF) interference, has been traditionally treated as a disturbance in the design of
wireless communications systems. However, RF signals carry information
as well as energy at  the  same  time. Such a dual nature of EM interference is
stimulating a significant interest  in communications systems  powered  by  harvested  ambient  energy. In particular,
{\em ambient backscatter} has emerged as a novel communication paradigm,
where a small passive device can
transmit its own data by backscattering the
EM/RF wave deriving from existing or {\em legacy} communication systems, such as
digital TV (DTV) broadcasting, cellular systems, or wireless
local area networks (LANs), e.g., Wi-Fi.
Unlike traditional backscatter systems, such as
radio frequency identification (RFID) ones \cite{Boy2014,Dar2015},
ambient backscatter does not require a dedicated reader, which
allows for direct device-to-device (D2D) and even multi-hop
communications.
Recently, this new communication paradigm has been receiving much attention
\cite{Liu,Kello,Bharadia,Ma2015,Lu2015,Wang2015},
since it can be embedded into inexpensive objects
in order to fulfil the ubiquitous and pervasive
communication vision of the Internet-of-Things (IoT) \cite{Sha}.

The main principles of ambient backscatter
were first introduced in \cite{Liu},
where also a simple prototype is developed,
which  harvests DTV energy to
achieve D2D communications with
rates of $1$ kbps over a range of about
$8$ m outdoor and $5$ m indoor.
In \cite{Kello}, the same principles are
exploited to allow a passive device or {\em tag}
to directly connect to the Internet
by leveraging on an existing Wi-Fi infrastructure.
In particular, in the scenario of \cite{Kello},
the tag can establish bidirectional communications with
a Wi-Fi device by modulating the channel state information (CSI) or
received signal strength indicator (RSSI) of the
Wi-Fi channel (in the uplink)
or by simple on-off modulation (in downlink),
achieving rates of $0.5$ kbps in uplink over a range
of $1$ m and up to $20$ kbps over $2.2$ m in downlink.
A significant improvement over this scheme
is the BackFi system proposed in \cite{Bharadia},
wherein backscatter communications
can achieve at least $1$ Mbps over a $5$m-range in uplink, by exploiting the signal cancellation
principles of full-duplex systems \cite{Ash}.

In \cite{Ma2015,Lu2015,Wang2015} the ambient backscatter approach  is extended to systems
where the backscatter receiver (called the reader)
is equipped with multiple antennas;
moreover, a detailed analysis of the system
from a signal processing perspective
is carried out, by assuming that the wireless channel obeys
a frequency-flat  block-fading model.
Since the tag employs low-rate differentially-encoded
on-off signaling, the reader
can decode its information by employing simple
noncoherent detection strategies.
The performance analysis of the approach proposed in \cite{Ma2015,Lu2015,Wang2015}  is
carried out in terms of bit-error rate (BER),
both analytically and by Monte Carlo simulations.

\IEEEpubidadjcol

Existing research on ambient backscatter has covered both
experimental and theoretical aspects.
However, to the best of our knowledge, an investigation of the ultimate
performance limits of ambient backscatter,
in terms of information-theoretic figures,
such as the ergodic or outage capacity, is still lacking.
We aim at filling this gap, by evaluating in this paper
the capacity (i.e., the maximum achievable transmission rate)
of ambient backscatter communications systems.
Our analysis assumes that the legacy system
employs a multicarrier modulation, which is ubiquitous
in modern communication systems, whereas the backscatter system
transmits at lower bit-rates by adopting simple single-carrier
techniques.
We evaluate typical information-theoretic figures of merit
for both the legacy and the backscatter systems,
by assuming a symbol variance constraint for the legacy system
and both symbol variance and amplitude constraints for the backscatter one.
Our results allow one to assess
the maximum data-rate achievable by the backscatter system,
and also show somewhat surprisingly that, since the backscatter transmitter acts as a relay
towards the legacy receiver, the legacy system can even benefit
of ambient backscatter, provided that some reasonable
assumptions are met. In other words, ambient backscatter is not only a viable
means of opportunistically capitalizing on the energy carried out
by RF signals, but it is also a way of
turning EM interference into a form of diversity.

The paper is organized as follows.
The system model is introduced in Section~\ref{sec:model}.
General assumptions underlying the performance
analysis are pointed out in Section~\ref{sec:assump}.
The analytical performance analysis
is carried out in Sections~\ref{sec:analysis-legacy} and
\ref{sec:analysis-back} for the legacy and the backscatter system, respectively.
Numerical results corroborating our analysis are reported in
Section \ref{sec:simul}. Conclusions are drawn in Section \ref{sec:concl}.

\begin{figure}[tbp]
\centering
\includegraphics[width=0.7\linewidth, trim = 0 0 0 0]{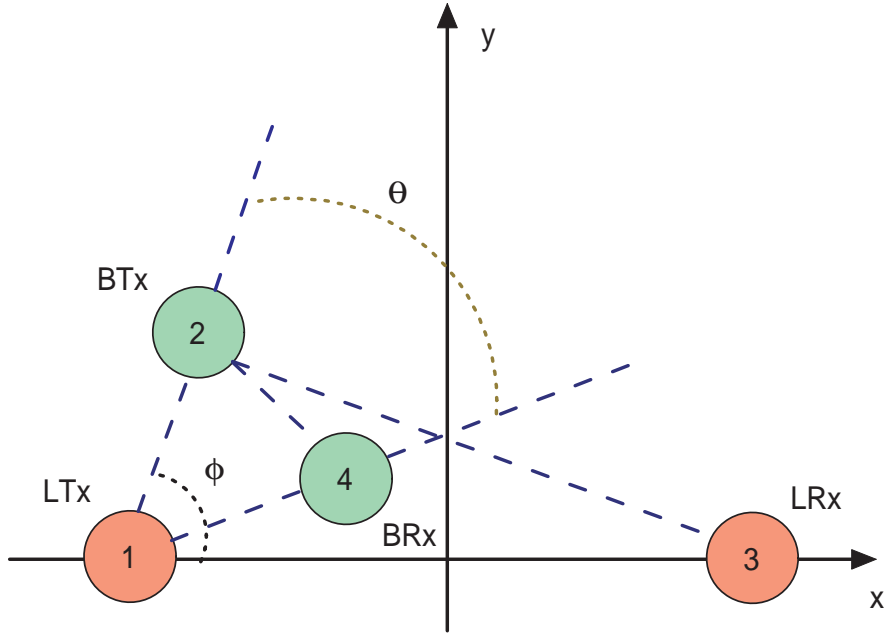}
\caption{
The considered wireless network model: in red,
the legacy transmitting (node $1$) and receiving (node $3$) devices;
in green, the backscatter transmitter (node $2$) and its
intended recipient (node $4$).
}
\label{fig:figure_1}
\end{figure}

\section{Ambient backscatter system model}
\label{sec:model}

In this section, we introduce a model for ambient backscatter
communications that harvest energy from legacy transmissions:
our model builds on the previous works \cite{Liu,Kello,Bharadia}.

The considered wireless network is depicted in Fig.~\ref{fig:figure_1}:
it is composed of a {\em legacy}\sfootnote{Hereinafter, similarly to \cite{Liu}, the
term ``legacy'' refers to existing wireless communications technologies, such as, e.g., DTV, cellular, and Wi-Fi systems.} transmitter-receiver (LTx/LRx) pair
and a
{\em backscatter} transmitter (BTx) that wishes to transmit information-bearing
symbols to an intended recipient (BRx).
In the sequel, the devices LTx, BTx, LRx, and BRx will be labelled as
nodes $1$, $2$, $3$, and $4$, respectively.
Specifically, the LTx and LRx are {\em active} devices, i.e., they have internal power sources to  modulate and demodulate, respectively, the relevant RF signals.
On the other hand, the BTx is a {\em passive} device, i.e., it does not include any active
RF component, and communicates
using only the power that it harvests from the RF signals transmitted by the LTx.
Finally, the BRx may be either passive or might use typical active RF
electronics to demodulate the signal backscattered by the BTx.

The LTx adopts a {\em multicarrier} modulation scheme with
$M$ subcarriers. The block of data
to be transmitted by the LTx within the $n$th ($n \in \Zset$)
frame of length $\Ts$ is denoted as
$\s(n) \eqdef [s^{(0)}(n), s^{(1)}(n), \ldots, s^{(M-1)}(n)]^\trasp \in \Cset^M$,
whose entries are independent and identically distributed (i.i.d.)
zero-mean circularly symmetric complex
symbols, with variance $\Ps \eqdef \Es[|s^{(m)}(n)|^2]$,
for any  $m \in \mathcal{M} \eqdef \{0,1,\ldots, M-1\}$ and
$n \in \Zset$. The vector $\s(n)$ is subject to conventional
multicarrier precoding, encompassing
$M$-point inverse discrete Fourier
transform (IDFT), followed by cyclic prefix (CP) insertion of length $\Lcp < M$.
It results that $\Ts \eqdef P \, \Tc$, with
$P \eqdef M + \Lcp$ and $\Tc$ denoting the
sampling period of the legacy system.
The data block transmitted by the LTx can be compactly
expressed \cite{Wang}
as
$\u(n) = \Tcp \, \Widft \, \s(n)$,
where
$\Tcp \eqdef [\I_{\text{cp}}^\trasp, \I_M]^\trasp \in \Rset^{P \times M}$,
with $\I_{\text{cp}} \in \Rset^{\Lcp \times M}$
obtained from $\I_M$ by picking its last $\Lcp$ rows,
and $\Widft \in \Cset^{M \times M}$ is
the unitary symmetric IDFT matrix \cite{Wang}.\sfootnote{Besides
standard notations, we adopt the following ones:
matrices [vectors] are denoted
with upper [lower] case boldface
letters (e.g., $\A$ or $\a$);
the superscripts
$*$, $\trasp$, $\herm$, and $-1$
denote the conjugate,
the transpose, the conjugate transpose, and
the inverse of a matrix, respectively;
$\log(\cdot)$ is taken to the base $2$;
the operator $\Es(\cdot)$ denotes ensemble averaging;
$\bm{O}_{m \times n} \in \Rset^{m \times n}$
and $\I_{m} \in \Rset^{m \times m}$
denote the null and
the identity matrices, respectively;
matrix $\A= \diag (a_{0}, a_{1}, \ldots, a_{n-1})$
is diagonal;
$\F \in \Rset^{n \times n}$ and $\B \in \Rset^{n \times n}$
denote the Toeplitz ``forward shift" and ``backward shift"
matrices \cite{Horn}, respectively,
where the first column of $\F$ and the first row
of $\B$ are given by $[0, 1, 0,  \ldots, 0]^T$
and $[0, 1, 0, \ldots, 0]$, respectively;
a circular symmetric complex Gaussian random vector $\bm{x} \in \Cset^n$
with mean $\bm{\mu} \in \Cset^n$ and covariance matrix $\bm{K} \in \Cset^{n
\times n}$ is denoted as $\bm{x} \sim {\cal CN}(\bm{\mu},\bm{K})$.
}
The entries of $\u(n)$ are subject to D/A plus RF conversion for transmission over the wireless channel.

On the other hand, due to  its power limitation,
the BTx transmits in a narrower
bandwidth  with respect to the legacy system (higher
data rates consume more power and energy).
Specifically, the BTx has a $Q$-order symbol sequence
$\{b(n)\}_{n \in \Zset} \in \mathcal{B} \eqdef \{\beta_1, \beta_2, \ldots, \beta_Q\}$ of i.i.d. zero-mean circularly symmetric complex
symbols destined for the BRx,
with variance $\Pb \eqdef \Es[|b(n)|^2]$,
for any $n \in \Zset$,
and signaling interval $T_s$.
Such a sequence is arranged in consecutive
frames of $B \in \Nset$ symbols, whose duration is less than or equal to
the coherence time $T_{\text{coh}} \eqdef B \, T_s$ of the
channels. 
It is noteworthy
that one symbol is transmitted by the BTx per each frame
of the legacy system.

\subsection{Signal backscattered by the BTx}
\label{sec:model-BTx}

Since the BTx is passive, it cannot
initiate transmissions on its own.
Once the LTx transmits the block $\u(n)$, the
EM wave propagates toward the BTx.
When the wave reaches the BTx, its antenna is excited and the RF power
is converted to direct current (DC) power through a {\em power harvester}. This DC
voltage is then able to power the control logic on the chip, whose
task is to modulate the reflected EM wave.

Regarding the  $1 \to 2$ link,
a frequency-selective and quasi-static channel model is assumed.
Specifically, during an interval of duration $T_{\text{coh}}$,
the channel impulse response spans $L_{12} \in \Nset$
sampling periods $T_c$;  hence,
the resulting discrete-time channel
$c_{12}(\ell)$ is a causal system of order $L_{12}$, i.e.,
$c_{12}(\ell) \equiv 0$ for $\ell \not \in \{0,1,\ldots, L_{12}\}$.
Moreover, the $1 \to 2$ link is
characterized by the (integer) time offset (TO)
$\theta_{12} \in \Nset$,
modeling the fact that the BTx does not know where the
multicarrier blocks of the legacy system start.\footnote{The fractional
TO is incorporated as part of $\{c_{12}(\ell)\}_{\ell=0}^{L_{12}}$.}
Finally, since the BTx simply remodulates the carrier of the LTx,
we assume in the sequel that the carrier frequency offset (CFO) is
negligible.\sfootnote{A CFO may occur as a
result of the Doppler effect from a mobile BTx, which is an unimportant
phenomenon in backscatter systems \cite{Liu,Kello,Bharadia}.}
Under these assumptions and provided that
$L_{12}+\theta_{12} \le P-1$,\sfootnote{\label{foot:noibi2}In general,
the received block within the $n$th frame is affected
not only by the IBI of the previous frame $n-1$
but also by the IBI of the $(n-2)$th frame. The
assumption $L_{ik}+\theta_{ik} \le P-1$
ensures that the sum of the TO and the
channel order turns out to be within one frame, such that
the $n$th received block
is impaired only by the IBI of the previous frame.
}
the baseband-equivalent block received by the BTx within the $n$th
frame can be written as
\be
\tilde{\r}_2(n) = \tilde{\C}_{12}^{(0)} \, \u(n)+ \tilde{\C}_{12}^{(1)} \, \u(n-1)
\label{eq:vetr2}
\ee
where $\tilde{\r}_2(n) \eqdef [r_2^{(0)}(n), r_2^{(1)}(n),
\ldots, r_2^{(P-1)}(n)]^\trasp \in \Cset^{P}$,
\barr
\tilde{\C}_{12}^{(0)} & \eqdef \sum_{\ell=0}^{L_{12}} c_{12}(\ell) \, \F^{\ell+ \theta_{12}} \in \Cset^{P \times P}
\label{eq:C120}
\\
\tilde{\C}_{12}^{(1)} & \eqdef \sum_{\ell=0}^{L_{12}} c_{12}(\ell) \, \B^{P-\ell- \theta_{12}} \in \Cset^{P \times P}
\label{eq:C121}
\earr
are Toeplitz lower- and upper-triangular matrices, respectively,
and we have neglected the noise
introduced by the BTx  \cite{Boy2014,Stock}, since the latter employs only passive components and does not
perform sophisticated signal processing operations.
It is worth noticing that the last $P-L_{12}-\theta_{12}$ rows of the matrix
$\tilde{\C}_{12}^{(1)}$ are identically zero, that is, the interblock interference (IBI)
contribution is entirely contained in the first $L_{12}+\theta_{12}$ entries
of the received vector $\tilde{\r}_2(n)$.

\begin{figure}
\centering
\includegraphics[width=0.7\linewidth, trim=0 10 0 10]{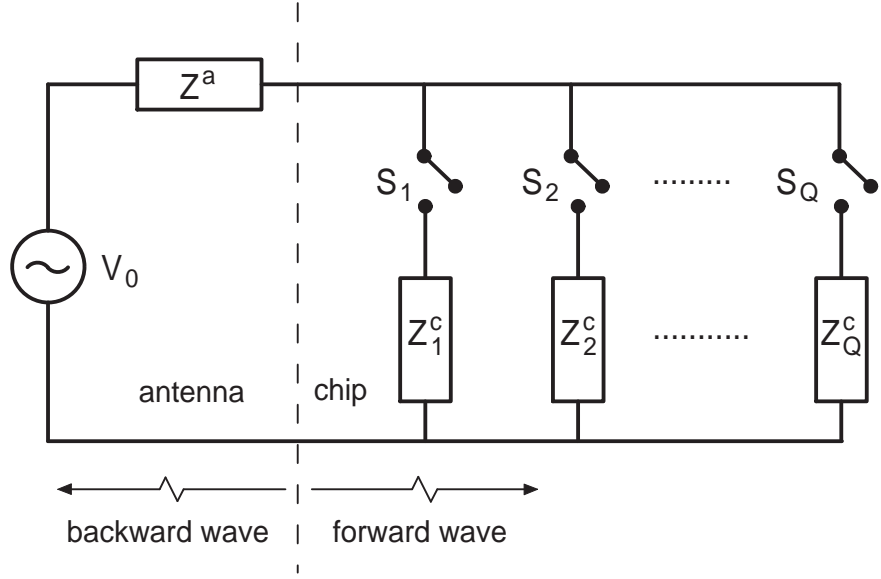}
\caption{Equivalent Th\'{e}venin circuit of the multilevel backscatter transmitter.}
\label{fig:fig_2}
\end{figure}

In our ambient backscatter framework, the BTx acts as a
digital multilevel modulator, mapping each information symbol onto a set of
$Q$ waveforms by means of a proper variation of its chip impedance \cite{Tho}.
To elaborate upon this point,
Fig.~\ref{fig:fig_2} reports the equivalent
Th\'{e}venin circuit \cite{King} of the BTx front-end,
where the sine wave generator $V_0$ models the
sinusoidal voltage induced by the power density of the incident EM field,
$Z^\text{a}=R^\text{a} +j \, X^\text{a}  \in \Cset$ is the
antenna impedance,
and $Z_{q}^\text{c}= R_{q}^\text{c}+ j \, X_{q}^\text{c} \in \Cset$
are $Q$ distinct values of the
BTx chip impedance, for $q \in \mathcal{Q} \eqdef \{1,2, \ldots, Q\}$.
The maximum power available from the
generator is given by  $\Pmax \eqdef |V_0|^2/(8 \, R^\text{a})$.

At the reference plane denoted by the dashed line in Fig.~\ref{fig:fig_2},
due to the  impedance discontinuity, two power waves are generated:
a {\em (nonreflecting) forward wave} propagating to the right
and a {\em (reflecting) backward wave} giving rise to the backscattered field.
When the switch $S_q$ is closed,
i.e., the chip impedance of the BTx takes on the value $Z_{q}^\text{c}$,
the average power harvested by the BTx is given \cite{Kuro} by
$\Phq = (1-|\Gamma_{q}|^2) \, \Pmax$
$(q \in \mathcal{Q})$,
where\sfootnote{The power wave reflection coefficient $\Gamma_{q}$ depends on the chip impedance that,  in its turn, depends on the chip input power.
A linearized model is herein assumed for the power wave reflection
coefficient \cite{Arn}, according to which $\Gamma_{q}$ does not depend on the incident power.}
\be
\Gamma_{q} = \frac{(Z^\text{a})^*-Z_{q}^\text{c}}{Z^\text{a}+Z_{q}^\text{c}}
\label{eq:refl_coeff}
\ee
is the {\em power wave reflection
coefficient} $\Gamma_{q} \in \Cset$.
The squared magnitude $0 \le |\Gamma_{q}|^2 \le 1$ of
the power wave reflection
coefficient is referred to as the {\em power reflection coefficient} \cite{Kuro}:
it measures the fraction of $\Pmax$ that is not delivered
to the chip of the BTx.
It is worth noticing that, if $(Z^\text{a})^*=Z_{q}^\text{c}$
({\em impedance matching condition}), then  $\Gamma_{q}=0$:
in this case, the tag achieves maximum average power harvesting
$\Pmax$ and, in theory, there is no backscattered field.
Hence, an impedance mismatch $(Z^\text{a})^* \neq Z_{q}^\text{c}$
is necessary to reflect part of the energy from the BTx antenna back
to the intended recipient BRx.

The symbol sequence $\{b(n)\}_{n \in \Zset}$ can be embedded in the
backscattered signal by carefully choosing the chip impedances
$Z_{1}^\text{c}, Z_{2}^\text{c}, \ldots, Z_{Q}^\text{c}$.
Each chip impedance in Fig.~\ref{fig:fig_2}
corresponds to a point of the
symbol constellation $\mathcal{B}$.
More precisely, to produce impedance values realizable with
passive components, all the power wave reflection coefficients
$\Gamma_1, \Gamma_2, \ldots, \Gamma_Q$ are confined
in the complex plane within a circle centered at the origin
with radius smaller than or equal to one. These coefficients
are then scaled by a constant $0 \le \alpha \le 1$ such that
\be
\Gamma_q = \alpha \, \beta_q
\quad (q \in \mathcal{Q})
\label{eq_gamma_symbol}
\ee
with $|\beta_q| \le 1$.
Eq.~\eqref{eq_gamma_symbol} establishes
a one-to-one mapping
between the information symbols of the BTx and
the power wave reflection coefficients of its chip.
Such a mapping is generally referred to
as {\em backscatter} or {\em load modulation} \cite{Tho}.
The choice of $\alpha$ governs the harvesting-performance
tradeoff of the backscatter communication process. Indeed, values of $\alpha$ closer
to one allows the BTx to reflect increasing amounts of the incident field back
to the BRx, resulting thus in greater backscatter signal strengths
(i.e., for a target symbol error probability at the BRx, larger communication ranges).
On the other hand, values of $\alpha$ much smaller than one
allow a larger part of the incident field to be absorbed by
the RF-to-DC conversion circuits of the BTx, hence improving power
conversion (i.e., $\Phq$) at the expense of backscatter signal strength.
We note that $\alpha=0$ accounts for the case when
the backscatter system is in sleep mode and, hence, only
the legacy transmission is active.

Once $\alpha$ and $\mathcal{B}$ have been chosen in accordance with certain criteria
\cite{Boy2014}
and, thus, the power wave reflection coefficients are identified through \eqref{eq_gamma_symbol}, the chip impedances
$Z_{1}^\text{c}, Z_{2}^\text{c}, \ldots, Z_{Q}^\text{c}$
corresponding to the designed
signal constellation can be obtained from \eqref{eq:refl_coeff}
as follows
\be
Z_{q}^\text{c} = \frac{(Z^\text{a})^*- Z^\text{a} \, \Gamma_{q} \, }{1+\Gamma_{q}}
\quad (q \in \mathcal{Q})
\label{eq:impedance}
\ee
where $Z^\text{a}$ is  a
given parameter. In practice, some constraints may be imposed on the
chip impedances \eqref{eq:impedance}: for instance, to use high-quality electronic components and/or  reduce the physical size of the BTx,  it might be required to use resistors and capacitors,
by hence eliminating inductors \cite{Tho}.

According to the antenna scatterer theorem \cite{Han},
the EM field backscattered
from the antenna of the BTx can be divided
\cite{Han} into load-dependent (or {\em antenna mode}) scattering and load-independent
(or {\em structural mode}) one:
the former component can be associated
with re-radiated power and depends on
the chip impedances of the BTx, whereas the latter one can be interpreted as
scattering from an open-circuited antenna.
Therefore,
with reference to antenna mode scattering and
accounting for \eqref{eq_gamma_symbol},
the $p$th  baseband-equivalent $T_c$-spaced sample backscattered by the
BTx during the $n$th frame of the legacy system assumes the expression
$x_{2}^{(p)}(n) =  \Gamma(n) \, r_{2}^{(p)}(n)$
$(p \in \mathcal{P})$, where
$\Gamma(n) \eqdef \alpha \, b(n)$ is
a discrete random variable assuming the values
$\Gamma_1, \Gamma_2, \ldots, \Gamma_Q$,
whereas $b(n) \in \mathcal{B}$ is the symbol transmitted
by the BTx during the $n$th frame.
The corresponding block model reads as
\be
\tilde{\x}_2(n)  =  \Gamma(n) \, \tilde{\r}_2(n)
= \alpha \, b(n) \, \tilde{\r}_2(n)
\label{eq:vetx2}
\ee
where $\tilde{\x}_2(n) \eqdef [x_2^{(0)}(n), x_2^{(1)}(n),
\ldots, x_2^{(P-1)}(n)]^\trasp \in \Cset^{P}$
and $\tilde{\r}_2(n)$ is given by \eqref{eq:vetr2}.

\subsection{Signal received by the LRx}
\label{sec:model-LRx}

With reference to the $1 \to 3$ and $2 \to 3$ links, we maintain the
same assumptions previously made for the $1 \to 2$ link:
basically,  for $i \in \{1,2\}$, within the coherence time $T_{\text{coh}}$,
the resulting discrete-time channel
$c_{i3}(\ell)$ is a causal system of order $L_{i3}$, i.e.,
$c_{i3}(\ell) \equiv 0$ for $\ell \not \in \{0,1,\ldots, L_{i3}\}$,
and $\theta_{i3} \in \Nset$ is the corresponding TO.
Since the BTx reflects the RF signal transmitted by the LTx,
both $1 \to 3$ and $2 \to 3$ transmissions
occur at the same RF frequency.
For such a reason, we assume that the corresponding CFOs
are equal and  can be accurately
estimated and compensated at the LRx through conventional
techniques \cite{Mengali_book}.

Provided that
$L_{13}+\theta_{13} \le P-1$ and
$L_{23}+\theta_{23} \le P-1$ (see footnote~\ref{foot:noibi2}),
accounting for \eqref{eq:vetr2} and
\eqref{eq:vetx2}, after CFO compensation,
the baseband-equivalent  vector received by the LRx
within the $n$th frame of the legacy system
can be expressed as
\begin{multline}
\tilde{\r}_3(n) =
\tilde{\C}_{13}^{(0)}\, \u(n)+ \tilde{\C}_{13}^{(1)} \, \u(n-1) +
\tilde{\C}_{23}^{(0)} \, \tilde{\x}_2(n) + \tilde{\C}_{23}^{(1)} \, \tilde{\x}_2(n-1) + \tilde{\v}_3(n)
\\
= \left[ \tilde{\C}_{13}^{(0)} + \alpha \, b(n) \,
\tilde{\C}_{23}^{(0)} \, \tilde{\C}_{12}^{(0)} \right] \u(n)
+
\left [\tilde{\C}_{13}^{(1)}  +
\alpha \, b(n) \, \tilde{\C}_{23}^{(0)} \,  \tilde{\C}_{12}^{(1)}
+ \alpha \,  b(n-1) \, \tilde{\C}_{23}^{(1)} \, \tilde{\C}_{12}^{(0)} \right] \u(n-1) + \tilde{\v}_3(n)
\label{eq:rtilde3}
\end{multline}
where $\{\tilde{\C}_{13}^{(0)},
\tilde{\C}_{13}^{(1)} \}$ and
$\{\tilde{\C}_{23}^{(0)},
\tilde{\C}_{23}^{(1)}\}$ can be obtained from
\eqref{eq:C120} and \eqref{eq:C121}
by replacing $\{L_{12}, c_{12}(\ell), \theta_{12}\}$
with $\{L_{13}, c_{13}(\ell), \theta_{13}\}$ and $\{L_{23}, c_{23}(\ell),
\theta_{23}\}$, respectively, and
$\tilde{\v}_3(n) \in \Cset^{P}$ accounts for the structural mode scattering, which
is independent of the BTx chip impedances, as well as for thermal noise.
We have also observed that
$\tilde{\C}_{23}^{(1)} \, \tilde{\C}_{12}^{(1)} =\Zero_{P \times P}$,
under the assumption that
\be
L_{12}+L_{23}+\theta_{12}+\theta_{23} \le P-1 \: .
\label{eq:dis}
\ee

The set of lower (upper) triangular Toeplitz matrices possesses an eminent algebraic structure: indeed, such a set is an {\em algebra} \cite{Horn}. In particular, the
product of any lower (upper) triangular Toeplitz matrices is a lower (upper) triangular Toeplitz matrix, too. Indeed, it is directly verified that, if \eqref{eq:dis} holds,
the product $\tilde{\C}_{23}^{(0)} \, \tilde{\C}_{12}^{(0)}$ is a
lower-triangular Toeplitz matrix
having as first column
$[\bm{0}_{\theta_{12}+\theta_{23}}^\trasp, \c_{123}^\trasp(n),
\bm{0}^\trasp_{P-L_{12}-L_{23}-\theta_{12}-\theta_{23}-1}]^\trasp$,
where the vector $\c_{123}\in \Cset^{L_{12}+L_{23}+1}$ collects the samples
of the (linear) convolution
between $\{c_{12}(\ell)\}_{\ell=0}^{L_{12}}$ and
$\{c_{23}(\ell)\}_{\ell=0}^{L_{23}}$.
Under the assumption that
\be
\Lcp \ge \max(L_{13}+\theta_{13}, L_{12}+L_{23}+\theta_{12}+\theta_{23})
\label{eq:CP}
\ee
the IBI contribution in \eqref{eq:rtilde3} can be completely discarded by dropping the first $\Lcp$ components of
$\tilde{\r}_3(n)$, since it is verified by direct inspection that:
(i) only the first $L_{12}+L_{23}+\theta_{12}+\theta_{23}$ rows
of $\tilde{\C}_{23}^{(0)} \,  \tilde{\C}_{12}^{(1)}$ are possibly nonzero;
(ii)
the last $P-L_{23}-\theta_{23}$ rows of the matrix $\tilde{\C}_{23}^{(1)}$ are identically zero and, hence,
the nonzero entries of $\tilde{\C}_{23}^{(1)} \, \tilde{\C}_{12}^{(0)}$
are located within its first $L_{23}+\theta_{23}$ rows;
(iii)
the last $P-L_{13}-\theta_{13}$ rows of
$\tilde{\C}_{13}^{(1)}$ are identically zero.
Therefore,  if \eqref{eq:CP} is fulfilled, after discarding the CP, performing
$M$-point discrete Fourier
transform (DFT),
the resulting frequency-domain data block
$\r_3(n) \in \Cset^{M}$ is given by
\be
\r_3(n) =\bCcal_3 \, \s(n) + \v_3(n)
\label{eq:r3}
\ee
where $\bCcal_3 \eqdef \diag[\Psi_3(0), \Psi_3(1), \ldots, \Psi_3(M-1)]$,
whose diagonal entries  are given by
\be
\Psi_3(m) \eqdef \Psi_{13}(m)+\alpha \, b(n) \, \Psi_{12}(m) \, \Psi_{23}(m)
\label{eq:C3-DFT}
\ee
for $m \in \mathcal{M}$, with
\be
\Psi_{ik}(m) \eqdef e^{-j \frac{2 \pi}{M} \theta_{ik} m}
\sum_{\ell=0}^{L_{ik}} c_{ik}(\ell) \, e^{-j \frac{2 \pi}{M} \ell m}
\label{eq:single-DFT}
\ee
and $\v_3(n) \in \Cset^{M}$ is obtained from $\tilde{\v}_3(n)$
by discarding its first $\Lcp$ entries and performing $M$-point DFT.

\vspace{3mm}
{\em Remark 1}: It is note\-worthy from \eqref{eq:rtilde3}-\eqref{eq:r3} that the
signal backscattered by the BTx may create additional paths
from the LTx to the LRx, which increases
multipath propagation on the legacy channel.
In particular, if $L_{12}+L_{23}+\theta_{12}+\theta_{23} >
L_{13}+\theta_{13}$, in accordance with \eqref{eq:CP},
such an additional multipath requires a corresponding increase of
the CP length in order to avoid  both IBI and intercarrier
interference (ICI) after CP removal, which may
worsen the performance of the legacy system.
In summary, the price to pay for allowing ambient backscatter
is an oversizing of the CP length, thus leading to an  inherent reduction of the transmission data rate of the legacy system. However, such a loss
turns out to be negligible if the number $M$
of subcarriers is significantly greater than $\Lcp$.
Most important, we show in Section~\ref{sec:analysis-legacy}
that,  if the legacy system is designed to
fulfil \eqref{eq:CP}, it might {\em even} achieve
a performance gain.

\vspace{3mm}
{\em Remark 2}: We note that assumption
\eqref{eq:CP} requires only upper bounds (rather than the
exact knowledge) on the channel orders and TOs. This is a
reasonable assumption in the considered
scenario. Indeed, in general, depending on the transmitted
signal parameters (carrier frequency and bandwidth) and environment
(indoor or outdoor), the maximum channel multipath
spread is known. For legacy systems,  particular
synchronization  policies are typically adopted
to drastically reduce the asynchronisms \cite{Morelli},
whereas, for ambient backscatter communications, the
distances among the LTx, BTx, and the BRx are very small.
Therefore, the TOs  are confined to a small uncertainty interval,
whose support can be typically predicted.

\subsection{Signal received by the BRx}
\label{sec:model-BRx}

Concerning the $1 \to 4$ and $2 \to 4$ links, we maintain the
same assumptions previously made for the
$1 \to 2$, $1 \to 3$, and $2 \to 3$ links:
in summary,  for $i \in \{1,2\}$, within the coherence time $T_{\text{coh}}$,
the resulting discrete-time channel
$c_{i4}(\ell)$ is a causal system of order $L_{i4}$, i.e.,
$c_{i4}(\ell) \equiv 0$ for $\ell \not \in \{0,1,\ldots, L_{i4}\}$,
and $\theta_{i4} \in \Nset$ is the corresponding TO.
Similarly to Subsection~\ref{sec:model-LRx}, we assume that
the $1 \to 4$ and $2 \to 4$ links have the same CFO,
which will be denoted as $\nu \in (-1/2,1/2)$ in the sequel
(it is normalized to the subcarrier spacing $1/T_c$).

Under the assumption that
$L_{14}+\theta_{14} \le P-1$ and
$L_{24}+\theta_{24} \le P-1$ (see footnote~\ref{foot:noibi2}),
the baseband-equivalent block received
by the BRx within the $n$th frame of the legacy system can be expressed as
shown at the top of this page in \eqref{eq:rtilde4},
\begin{figure*}[!t]
\normalsize
\begin{multline}
\tilde{\r}_4(n) = e^{j \frac{2 \pi}{M} \nu n P} \, \bSigma_{\nu} \left [
\tilde{\C}_{14}^{(0)} \, \u(n)+ \tilde{\C}_{14}^{(1)} \, \u(n-1) +
\tilde{\C}_{24}^{(0)} \, \tilde{\x}_2(n) + \tilde{\C}_{24}^{(1)} \, \tilde{\x}_2(n-1) \right]+ \tilde{\v}_4(n)
\\
= \alpha \, e^{j \frac{2 \pi}{M} \nu n P} \, \bSigma_{\nu} \left[ \tilde{\C}_{24}^{(0)} \, \tilde{\C}_{12}^{(0)}
\,  \u(n) + \tilde{\C}_{24}^{(0)} \,  \tilde{\C}_{12}^{(1)}  \, \u(n-1) \right] b(n)
\\ +
\alpha \, e^{j \frac{2 \pi}{M} \nu n P} \, \bSigma_{\nu} \,
\tilde{\C}_{24}^{(1)} \, \tilde{\C}_{12}^{(0)} \,  \u(n-1) \,
b(n-1)
\\ + e^{j \frac{2 \pi}{M} \nu n P} \, \bSigma_{\nu} \left[
\tilde{\C}_{14}^{(0)} \, \u(n)+ \tilde{\C}_{14}^{(1)} \, \u(n-1) \right]
+ \tilde{\v}_4(n)
\label{eq:rtilde4}
\end{multline}
\hrulefill
\end{figure*}
where
$\{\tilde{\C}_{14}^{(0)},
\tilde{\C}_{14}^{(1)} \}$ and
$\{\tilde{\C}_{24}^{(0)},
\tilde{\C}_{24}^{(1)}\}$ can be obtained from
\eqref{eq:C120} and \eqref{eq:C121}
by replacing $\{L_{12}, c_{12}(\ell), \theta_{12}\}$
with $\{L_{14}, c_{14}(\ell), \theta_{14}\}$ and $\{L_{24}, c_{24}(\ell),
\theta_{24}\}$, respectively, we have defined the diagonal matrix
$\bSigma_{\nu} \eqdef \diag[1, e^{j \frac{2 \pi}{M} \nu},
\ldots, e^{j \frac{2 \pi}{M} \nu (P-1)}] \in \Cset^{P \times P}$,
and
$\tilde{\v}_4(n) \Cset^{P}$ accounts for both the structural mode scattering
and thermal noise.

\vspace{3mm}
{\em Remark 3}: It is important to notice from \eqref{eq:rtilde4} that
the BRx experiences {\em frequency-selective fast fading}, since:
(i) the received signal is corrupted by the intersymbol interference (ISI)
of the previous symbol $b(n-1)$;
(ii)
the channel tap seen by the BRx varies with time
from sampling period to sampling period, due to
its dependence on the data $\{u^{(p)}(n)\}_{p=0}^{P-1}$ transmitted by the LTx,
and such a variation is $P$-times faster than
the symbol rate $1/T_s$ of the backscatter system.

\vspace{3mm}
Interestingly, by observing that
the nonzero entries of $\bSigma_{\nu} \, \tilde{\C}_{24}^{(1)}\, \tilde{\C}_{12}^{(0)}$
are located within its first $L_{24}+\theta_{24}$ rows
and the last $P-L_{14}-\theta_{14}$ rows of
$\bSigma_{\nu} \, \tilde{\C}_{14}^{(1)}$ are identically zero,
the BRx can resort to a simple
detection technique to completely remove its own ISI
and partially mitigate the interference generated by
the legacy transmission. More specifically, this
can be obtained by dropping the first
\be
L_b \ge \max\left(L_{14}+\theta_{14}, L_{24}+\theta_{24} \right)
\label{eq:lb}
\ee
components of $\tilde{\r}_4(n)$.
This operation is accomplished
by defining the matrix $\R_{b} \eqdef [\Zero_{N \times L_b}, \I_{N}] \in \Rset^{N \times P}$, with $N \eqdef P-L_b>0$, and
forming at the receiver the product $\R_{b} \, \tilde{\r}_4(n)$.
So doing, one has
\be
\R_{b} \, \tilde{\r}_4(n) = \alpha \, \c_4 \, b(n) + \d_4(n)
\label{eq:r4}
\ee
with
\barr
\c_4  & \eqdef
e^{j \frac{2 \pi}{M} \nu n P} \, \R_b \, \bSigma_{\nu} \left[ \tilde{\C}_{24}^{(0)} \, \tilde{\C}_{12}^{(0)} \,  \u(n)
+ \tilde{\C}_{24}^{(0)} \,  \tilde{\C}_{12}^{(1)} \, \u(n-1)\right] \in
\Cset^{N}
\label{eq:c4}
\\
\d_4(n) & \eqdef
e^{j \frac{2 \pi}{M} \nu n P} \, \R_b \, \bSigma_{\nu} \,
\tilde{\C}_{14}^{(0)} \, \u(n) + \R_b \, \tilde{\v}_4(n) \in \Cset^{N}
\label{eq:d4}
\earr
where it results that
$\R_b \, \bSigma_{\nu} \, \tilde{\C}_{24}^{(1)} \, \tilde{\C}_{12}^{(0)}=\Zero_{N \times P}$
and  $\R_b \, \bSigma_{\nu} \, \tilde{\C}_{14}^{(1)}=\Zero_{N \times P}$.
To fulfil \eqref{eq:lb}, some {\em a priori} knowledge is required at the BRx, which
can be acquired in practice (see Remark~2):
as explained in Section~\ref{sec:analysis-back}, such a knowledge at the BRx
depend on whether the BRx and LTx are spatially-separated nodes \cite{Liu}
or they are the co-located  \cite{Kello,Bharadia}, 
i.e., they are on the same machine.

\section{General assumptions for the analytical performance analysis}
\label{sec:assump}

The goal of the forthcoming Sections \ref{sec:analysis-legacy} and
\ref{sec:analysis-back} is twofold.
First, we aim at showing in Section \ref{sec:analysis-legacy} what is the influence of
the backscatter communication on the
achievable rates of the legacy system,
by assuming that the CP is long enough,
i.e., inequality \eqref{eq:CP} is fulfilled.
Second, under assumption \eqref{eq:lb},
we highlight in Section \ref{sec:analysis-back} what are
the ultimate rates of the backscatter communication,
by considering either the case when the nodes BRx and LTx
are co-located \cite{Kello,Bharadia}
or the situation in which they are 
spatially-separated nodes \cite{Liu}.
General assumptions are
reported in the sequel.

For $i \in \{1,2\}$ and $k \in \{2,3,4\}$,
with $i \neq k$,
the channel samples
$c_{ik}(0), c_{ik}(1), \ldots, c_{ik}(L_{i k})$
(encompassing  the  physical  channel  as  well  as  the
transmit/receive filters) are modeled as
i.i.d. zero-mean circularly symmetric complex
Gaussian random coefficients ({\em Rayleigh  fading model}),\sfootnote{Although
the  transmit/receive  filters  might
introduce statistical correlation among channel taps,
it is a common practice \cite{Morelli} to neglect such a correlation when
evaluating  the  performance  of  multicarrier  systems.
}
which are constant within the
coherence time $T_{\text{coh}}$, but
are allowed to vary independently
in different coherence intervals;
the variance $\Es[|c_{ik}(\ell)|^2] \eqdef \sigma_{ik}^2/(L_{ik}+1)$
of the $i \to k$ link  depends on the corresponding average
path loss.
Fading coefficients of different links are
statistically independent among themselves,
i.e.,  $c_{i_1k_1}(\ell)$ is statistically independent
of $c_{i_2k_2}(\ell)$ for $i_1 \neq i_2$ or
$k_1 \neq k_2$.

Since $c_{ik}(\ell)$ is a circularly symmetric complex Gaussian random variable
by assumption, then $c_{ik}(\ell)$ and $c_{ik}(\ell) \, e^{-j \frac{2 \pi}{M} (\ell+\theta_{ik}) m}$ have the same probability distribution \cite{Picinbono}, i.e.,
$c_{ik}(\ell) \, e^{-j \frac{2 \pi}{M} (\ell+\theta_{ik}) m} \sim \CN[0,\sigma_{ik}^2/(L_{ik}+1)]$,
for any $\ell$, $m$, and $n$. Consequently, one has
$\Psi_{ik}(m) \sim \CN(0, \sigma_{ik}^2)$.
It is seen from \eqref{eq:single-DFT} that,
even if the time-domain channel taps $\{c_{ik}(\ell)\}_{\ell=0}^{L_{ik}}$ are assumed to be uncorrelated, the corresponding DFT
samples $\Psi_{ik}(m_1)$ and $\Psi_{ik}(m_2)$
turn out to be correlated, for $m_1 \neq m_2 \in \mathcal{M}$.
%, i.e.,
%\begin{multline}
%\psi_{ik}(m_1,m_2) \eqdef \Es[\Psi_{ik}(m_1) \, \Psi_{ik}^*(m_2)]
%\\ =
%\frac{\sigma_{ik}^2}{L_{ik}+1} \, e^{-j \frac{2 \pi}{M} \theta_{ik} (m_1-m_2)} \,
%\Dir_{L_{ik}+1}\left(\frac{m_1-m_2}{M}\right)
%\end{multline}
%where, for $x \in \Rset$, we have defined the function
%\be
%\Dir_{L_{ik}+1}(x) \eqdef \frac{\sin[\pi (L_{ik}+1) x]}{\sin(\pi x)} \,
%e^{-j \pi L_{ik} x} \: .
%\ee
For $k \in \{3, 4\}$,
we assume that $\tilde{\v}_k(n) \sim \CN(\bm{0}_{P}, \sigma_{v_k}^2 \, \I_{P})$
with $\Es[\tilde{\v}_k(n_1) \, \tilde{\v}_k^\herm(n_2)]=\Zero_{P \times P}$,
for $n_1 \neq n_2 \in \Zset$.

Finally, channel coefficients,
information-bearing symbols, and noise samples
are all modeled as statistically independent random variables.

\section{Capacity analysis of the legacy system}
\label{sec:analysis-legacy}

Since the detection process at the LRx is
carried out on a frame-by-frame basis, we
omit the dependence on the frame index $n$ hereinafter.
Under the assumption that the realization $\bXi_3$ of $\bCcal_3$ is known at the LRx
(but not at the LTx), the channel output of  \eqref{eq:r3}
is the pair $(\r_3, \bCcal_3)$. Therefore, the {\em (coherent) ergodic (or Shannon) capacity}
of \eqref{eq:r3} is defined as (see, e.g., \cite{Telatar})
\be
\Cap_3 \eqdef  \sup_{f(\s) \in \mathcal{I}_{s}} \frac{\MI(\s; \r_3, \bCcal_3)}{M}
\quad \text{(in b/s/Hz)}
\ee
where $f(\s)$ is the probability density function (pdf) of $\s$,
$\mathcal{I}_{s}$ is the set of admissible input distributions
having the variance constraint $\Es(\|\s\|^2)=M \sigma_s^2$
and $\MI(\s; \r_3, \bCcal_3)$ denotes the mutual information \cite{Gallager_book,CoverThomas_book}
between  $\s$ and $(\r_3, \bCcal_3)$.
The ergodic capacity can be achieved if the length of
the codebook is long enough to reflect the
ergodic nature of fading  \cite{Biglieri}
(i.e.,  the duration of each transmitted codeword is much greater than the
channel coherence time).

By using the chain rule for mutual information
\cite{Gallager_book, CoverThomas_book} and observing that
$\s$ and $\bCcal_3$ are statistically independent, it results
that
$\MI(\s; \r_3, \bCcal_3)  =\MI(\s; \bCcal_3)
+\MI(\s; \r_3 \,|\, \bCcal_3)=\MI(\s; \r_3 \,|\, \bCcal_3) =
\Es_{\bCcal_3}[\MI(\s; \r_3 \,|\, \bCcal_3=\bXi_3)]$,
where $\MI(\s; \r_3 \,|\, \bCcal_3)$
is the mutual information
between  $\s$ and $\r_3$, given $\bCcal_3$.
It is shown in \cite{Telatar}
that, given $\bCcal_3=\bXi_3$, the input distribution that maximizes
$\MI(\s; \r_3 \,|\, \bCcal_3=\bXi_3)$ is
$\s \sim \CN(\bm{0}_{M}, \Ps \, \I_{M})$ and the corresponding
maximal mutual information $\MI_{\text{max}}(\s; \r_3 \,|\, \bCcal_3=\bXi_3)$
is given by
\be
\MI_{\text{max}}(\s; \r_3 \,|\, \bCcal_3=\bXi_3)  =  \log  \det \left(\I_M + \frac{\Ps}{\sigma_{v_3}^2} \, \bXi_3 \, \bXi_3^\herm \right)  \: .
\label{eq:cap-alea}
\ee
Consequently, one has
\be
\Cap_3  = \frac{1}{M} \, \Es \left[\log \det \left(\I_M + \frac{\Ps}{\sigma_{v_3}^2} \, \bCcal_3 \, \bCcal_3^\herm \right) \right] =
\frac{1}{M} \sum_{m=0}^{M-1}
\Es \left[\log \left( 1 + \frac{\sigma_s^2}
{\sigma_{v_3}^2} |\Psi_3(m)|^2 \right)\right]
\label{eq:C3-1}
\ee
where  $\Psi_3(m) $ has been defined in \eqref{eq:C3-DFT}.

A first step towards the analytical computation of $\Cap_3$
consists of observing
that, conditioned on the product $b \, \Psi_{12}(m)$,
$|\Psi_3(m)|^2$ turns out to be exponentially distributed
with mean
$\sigma_{13}^2 + \alpha^2 \, \sigma_{23}^2  \, |b|^2 \, |\Psi_{12}(m)|^2$
($m \in \mathcal{M}$).
Thus, by applying the conditional expectation rule \cite{Papoulis}, one obtains
\be
\Cap_3 = - \frac{\log e}{M} \sum_{m=0}^{M-1} \Es\left\{e^{{1}/{\Upsilon_3(m)}} \, \mathrm{Ei}\left[-{1}/{\Upsilon_3(m)}\right] \right\}
\label{eq:C3-2}
\ee
where
$\text{Ei}(x) \eqdef  \int_{-\infty}^{x} {e^{u}}/{u} \,
\mathrm{d}u$ denotes the exponential integral function, for $x < 0$, and
\be
\Upsilon_3(m) \eqdef \Gamma_{13} \, \left[1+ \alpha^2 \, \frac{\sigma^2_{23}}{\sigma^2_{13}} \, |b|^2 \, |\Psi_{12}(m)|^2 \right]
\label{eq:asnr3}
\ee
with $\Gamma_{13} \eqdef (\sigma_{13}^2 \, \sigma_s^2)/
{\sigma_{v_3}^2}$ representing the average signal-to-noise ratio (SNR)
over the  $1 \to 3$ link. When the backscatter system is inactive,
i.e., $\alpha=0$, in accordance with \cite{Ozarow}, the ergodic
capacity of the legacy system is  given by
\be
\Cap_3 |_{\alpha=0} = - e^{{1}/{\Gamma_{13}}} \,
\mathrm{Ei}\left(-{1}/{\Gamma_{13}}\right) \, \log e  \: .
\label{eq:C3-direct}
\ee
A first result can be obtained by comparing \eqref{eq:C3-2} and \eqref{eq:C3-direct}.
Indeed, since $\Upsilon_3(m) \ge \Gamma_{13}$ for any realizations
of $|b|^2$ and $|\Psi_{12}(m)|^2$ and, moreover,
$-e^{1/x} \, \text{Ei}(-1/x)$ is a monotonically
increasing function of $x \ge 0$, it follows that $\Cap_3 \ge \Cap_3 |_{\alpha=0}$.

\vspace{3mm}
{\em Remark 4}: If the constraint \eqref{eq:CP} on the CP length is satisfied, then   backscatter communications can even increase the ergodic capacity of
the legacy system. Strictly speaking, the interference
generated by the backscatter communication is turned
into a form of diversity for the legacy system.

\vspace{3mm}
To assess the performance gain $\Delta \Cap_3 \eqdef \Cap_3 -\Cap_3 |_{\alpha=0}$,
we use asymptotic expressions for $\Cap_3$ by considering both
low- and high-SNR regimes. With this goal in mind, we assume that
$\sigma_{i k}^2 = d_{i k}^{-\eta}$,
where $d_{i k}$ is the distance between
nodes $i$ and $k$, and
$\eta$ denotes the path-loss exponent.
Specifically, since
$-e^{1/x} \, \text{Ei}(-1/x) \to x$ as $x \to 0$ \cite{Ozarow}, 
in the low-SNR regime, i.e.,  
$\SNRL \eqdef \sigma_s^2/\sigma_{v_3}^2 \to 0$, one has
\be
\Cap_3 |_{\alpha=0} \to \Gamma_{13} \, \log e
\ee
and
\be
\Cap_3  \to \frac{\log e}{M} \sum_{m=0}^{M-1} \Es[\Upsilon_3(m)]
=
\Gamma_{13} \left[1+ \alpha^2 \, \sigma_b^2 \, \frac{\sigma^2_{12} \, \sigma^2_{23}}{\sigma^2_{13}}\right] \log e \: ,
\quad \text{for $\SNRL \to 0$}
\ee
which leads to
\be
\Delta \Cap_3 \to \alpha^2 \, \sigma_b^2 \,
\frac{\sigma_s^2}{\sigma_{v_3}^2} \,  \left(\frac{1}{d_{12} \, d_{23}}\right)^\eta \log e \: ,
\quad \text{for $\SNRL \to 0$} \: .
\label{eq:lowsnr}
\ee
where, according to \eqref{eq_gamma_symbol}, it results
that $\sigma_b^2=\Es(|b|^2) \le 1$.

On the other hand, by using the fact that
$-e^{1/x} \, \text{Ei}(-1/x) \to \log(1+x)-\gamma$ as $x \to + \infty$ \cite{Ozarow},
where $\gamma \eqdef \lim_{i \to \infty}\left[ i^{-1} \sum_{k=1}^{i} k^{-1} - \log(i) \right] \approx  0.57721$
is the Euler-Mascheroni constant, we have that,
in the high-SNR regime, i.e., when $\SNRL \to +\infty$,
\be
\Cap_3 |_{\alpha=0} \to [\log(1+\Gamma_{13})-\gamma] \log e
\ee
and,
moreover,
\be
\Cap_3 \to \frac{\log e}{M} \sum_{m=0}^{M-1} \Es\{\log[1+\Upsilon_3(m)]-\gamma\} \: ,
\quad \text{for $\SNRL \to +\infty$} \: .
\label{eq:C3-3}
\ee
To analytically compute the ensemble average in \eqref{eq:C3-3}, we
assume that the backscatter system employs
a constant-modulus constellation, e.g., $Q$-ary phase-shift keying (PSK),
with average energy $\sigma_b^2=1$. Henceforth,
$|b|=1$ and,
by observing that $|\Psi_{12}(m)|^2$ is exponentially distributed
with mean $\sigma_{12}^2$, after
some calculations, one has
\be
\Delta \Cap_3 \to - e^{{1}/{\Omega_{3}}} \,
\mathrm{Ei}\left(-{1}/{\Omega_{3}}\right) \, \log^2 e \: ,
\quad \text{for $\SNRL \to +\infty$}
\label{eq:highsnr}
\ee
with
\be
\Omega_3 \eqdef \alpha^2 \, \frac{\sigma_{23}^2 \, \sigma_{12}^2}{\sigma_{13}^2} =
\alpha^2 \, \left( \frac{d_{13}}{d_{12} \, d_{23}} \right)^\eta
\label{eq:omega3}
\ee
where
we observed that $\Gamma_{13}/(1+\Gamma_{13}) \to 1$
as $\SNRL \to +\infty$.
Two remarks are now in order.

\vspace{3mm}
{\em Remark 5}:
The capacity gain $\Delta \Cap_3$ increases
with $\alpha^2$, that is,
the greater the backscatter signal strength, the greater
the capacity gain of the legacy system.
Such a result directly comes from the fact that the
backscatter device can be regarded as a non-regenerative
relay for the legacy system.

\vspace{3mm}
{\em Remark 6}:
With reference to Fig.~\ref{fig:figure_1}, 
let the angle $\phi$ between nodes $2$ and $3$
and the distance $d_{13}$ between the
LTx and the LRx be fixed. 
As a consequence of the Carnot's cosine law
$d_{23} = (d_{12}^2+d_{13}^2 - 2 \, d_{12} \, d_{13} \, \cos\phi)^{1/2}$,
which can be  substituted in \eqref{eq:lowsnr} and
\eqref{eq:omega3}. By using standard calculus concepts, 
it can be verified that, in both
low- and high-SNR regimes, $\Delta \Cap_3$
is {\em not} a monotonic function of the distance $d_{12}$
between the LTx and the BTx, for each $\phi \in [0, 2 \pi)$. Indeed, it results that $\Delta \Cap_3$ is
a strictly decreasing function of $d_{12}/d_{13}$ when
$9 \, \cos^2 \phi - 8 < 0$, i.e., the angle $\phi$ belongs to the set
\begin{multline}
\mathcal{A} \eqdef \left \{\arccos(2 \, \sqrt{2}/3)  < a <
\pi-\arccos(2 \, \sqrt{2}/3) \, \text{and} \,
\pi +\arccos(2 \, \sqrt{2}/3)  < a <
2 \pi-\arccos(2 \, \sqrt{2}/3) \right \}
\end{multline}
i.e., the capacity gain decreases as the BTx moves away from the LTx.
On the other hand, when
$9 \,  \cos^2 \phi - 8 \ge  0$, i.e., $\phi \not \in \mathcal{A}$,
the capacity $\Delta \Cap_3$ monotonically increases for
$d_{\text{min}}(\phi) \le {d_{12}}/{d_{13}} \le d_{\text{max}}(\phi)$,
with
\barr
d_{\text{min}}(\phi) & \eqdef
\max \left(0,\frac{3 \, \cos \phi - \sqrt{9 \, \cos^2 \phi -8}}{4} \right)
\\
d_{\text{max}}(\phi) & \eqdef \max \left (0,
\frac{3 \, \cos \phi + \sqrt{9 \, \cos^2 \phi -8}}{4}\right )
\earr
otherwise, it monotonically decreases.
For instance, if
LTx, BTx, and LRx lie on the same line, i.e., $\phi=0$, the function
$\Delta \Cap_3$ monotonically decreases for $0 < d_{12}/d_{13} \le 1/2$ and $d_{12}/d_{13} >1$, while it increases when
$1/2 < d_{12}/d_{13}< 1$.
In this case,
the capacity gain of the legacy system increases
as the BTx gets closer and closer to either the LTx or the LRx.

\vspace{3mm}
If no  significant  channel  variability  occurs  during
the whole legacy transmission (i.e.,  the transmission duration of the
codeword is comparable to the
channel coherence time), a capacity  in  the  ergodic
sense does not exist.  In this case, the concept of {\em capacity
versus outage} has to be used \cite{Biglieri, Ozarow}.
Assume that codewords extend over a single legacy frame
and let the LTx encode data at a rate of $R_s$ b/s/Hz,
the outage probability of the legacy system is defined as
\be
{\mathsf{P}}_{\text{out},3} \eqdef
P\left\{ \frac{1}{M} \sum_{m=0}^{M-1}
\log \left[ 1 + \frac{\sigma_s^2}{\sigma_{v_3}^2} |\Psi_3(m)|^2 \right] < R_s \right\} \:.
\label{eq:pout3}
\ee
However, for the problem at hand, ${\mathsf{P}}_{\text{out},3}$ is
hard to compute analytically and does not lead to easily
interpretable results. Therefore,  we resort
to numerical simulations presented in Section~\ref{sec:simul}
to show the influence of the main system parameters
on the outage probability of the legacy communication.

\section{Capacity analysis of the backscatter system}
\label{sec:analysis-back}

In the subsequent analysis, we separately consider two
different network configurations. In the former case,
we focus on the scenario where the intended
recipient of the backscatter communication BRx and the legacy
transmitter LTx are co-located,
which is the situation considered in \cite{Kello,Bharadia}.
In the latter case, we study
the scenario where the BRx and the LTx are 
spatially-separated nodes, which is the situation considered in
\cite{Liu}.

For simplicity, we remove the IBI in \eqref{eq:c4} by
replacing condition \eqref{eq:lb} with the more restrictive one
\be
L_b \ge \max\left(L_{14}+\theta_{14}, L_{24}+\theta_{24},
L_{12}+L_{24}+\theta_{12}+\theta_{24} \right) \: .
\label{eq:lbb}
\ee
Under the assumption that
\be
L_{12}+L_{24}+\theta_{12}+\theta_{24} \le P-1
\label{eq:disss}
\ee
since
only the first $L_{12}+L_{24}+\theta_{12}+\theta_{24}$ rows
of $\tilde{\C}_{24}^{(0)} \,  \tilde{\C}_{12}^{(1)}$ might
not be zero, one thus has
$\R_b \, \bSigma_{\nu} \, \tilde{\C}_{24}^{(0)} \, \tilde{\C}_{12}^{(1)}
=\Zero_{N \times P}$. Obviously, removing the IBI in \eqref{eq:c4}
is not the best choice, since it does not allow one to exploit
the entire channel energy. However, such a contribution becomes
negligible for large values of $N$ (i.e., $P$). Moreover,
we assume herein that, in both the aforementioned
cases, the number of samples $L_b$ discarded from the received
backscatter signal \eqref{eq:rtilde4} is just equal to $\Lcp$.
We note that,  when $\Lcp=L_b$, then $N=M$ in \eqref{eq:r4}--\eqref{eq:d4}.
In this case, if \eqref{eq:dis} holds,
the product $\tilde{\C}_{24}^{(0)} \, \tilde{\C}_{12}^{(0)}$ is a
lower-triangular Toeplitz matrix
having as first column
$[\bm{0}_{\theta_{12}+\theta_{24}}^\trasp, \c_{124}^\trasp,
\bm{0}^\trasp_{P-L_{12}-L_{24}-\theta_{12}-\theta_{24}-1}]^\trasp$,
where the vector $\c_{124} \in \Cset^{L_{12}+L_{24}+1}$ collects the samples
of the (linear) convolution
between $\{c_{12}(\ell)\}_{\ell=0}^{L_{12}}$ and
$\{c_{24}(\ell)\}_{\ell=0}^{L_{24}}$.
This implies that the CP of the legacy system has to be designed
to satisfy both inequalities \eqref{eq:CP} and \eqref{eq:lbb}.
We would like to point out that, even though such an assumption
is made only to keep the analysis relatively simple from a mathematical
viewpoint, it is quite reasonable for small area networks.

\subsection{The BRx and LTx are co-located}
\label{sec:coincide}

When the intended recipient of the backscatter signal
and the legacy transmitter are co-located,  
the reference signal model can be obtained from \eqref{eq:r4}--\eqref{eq:d4} by replacing the subscript $4$ with $1$
and setting $\nu=0$, which implies that $\bSigma_{\nu}=\I_P$. In this
case, the matrix $\tilde{\C}_{11}^{(0)}$ models a self-interference channel
and $\tilde{\C}_{11}^{(0)} \, \u(n)$
represents direct leakage between the LTx transmit/receive chains
and/or reflections by other objects in the environment \cite{Bharadia}.

It is worth observing that the symbol vector $\s(n)$ [and, thus, $\u(n)$] is perfectly known at the LTx,
whereas the parameters $\{\c_{121}, \theta_{12}+\theta_{21}\}$, which uniquely identify
the matrix $\tilde{\C}_{21}^{(0)} \, \tilde{\C}_{12}^{(0)} $,
and $\{\c_{11}, \theta_{11}\}$, which uniquely identify the matrix $\tilde{\C}_{11}^{(0)}$, with
$\c_{11} \eqdef [c_{11}(0),
c_{11}(1), \ldots, c_{11}(L_{11})]^\trasp \in \Cset^{L_{11}+1}$,
can be estimated by allowing the insertion of training data
within each packet of $B$ symbols transmitted by the BTx.
More precisely, the self-interference parameters $\{\c_{11}, \theta_{11}\}$
can be estimated when there is no backscatter transmission:
this can be obtained at the protocol level by employing a {\em silent period}
of few symbols at the beginning of the packet \cite{Bharadia}, during which
the BTx does not backscatter (i.e.,  $\alpha=0$).
Once $\c_{11}$ and $\theta_{11}$ have been estimated by means of standard techniques
\cite{Mengali}, the self-interference contribution
can be subtracted from \eqref{eq:r4}.
After the silent period, the BTx modulates  training symbols on the backscatter
signal \cite{Bharadia}, which can be used to estimate
$\{\c_{121}, \theta_{12}+\theta_{21}\}$ through conventional methods \cite{Mengali}.

After performing the DFT, one gets
\be
\r_1(n)  \eqdef \Wdft \, \R_b \left[\tilde{\r}_1(n)-\tilde{\C}_{11}^{(0)} \, \u(n)\right]
= \alpha \, \bpsi(n) \, b(n) + {\v}_1(n)
\label{eq:r44}
\ee
where $\Wdft \eqdef \Widft^{-1}=\Widft^\herm$
defines the unitary symmetric DFT matrix \cite{Wang}
and the nonzero entries of the diagonal matrix
\be
\bCcal_{ik} \eqdef \diag[\Psi_{ik}(0), \Psi_{ik}(1), \ldots, \Psi_{ik}(M-1)]
\label{eq:diag}
\ee
are given by \eqref{eq:single-DFT},
$\bpsi(n) \eqdef \bpsi(n) \in \Cset^M$,
and $\v_1(n) \eqdef \Wdft \, \R_b \, \tilde{\v}_1(n) \in \Cset^{M}$.
On the basis of the above discussion, the vector $\bpsi(n)$
is assumed to be known at the LRx
and, thus, {\em coherent} receiving rules can be adopted at the LRx.
Moreover, we will omit the dependence on the frame index $n$ hereinafter.

According to \eqref{eq:r44}, given $\bpsi$,
a sufficient statistic for detecting $b$ from $\r_1$ is given by the scalar
\be
z_1\eqdef \bpsi^\herm \, \r_1 =
\alpha \, \|\bpsi \|^2 \, b +
\bpsi^\herm \,{\v}_1 \: .
\label{eq:z1}
\ee
Since sufficient statistics preserve mutual information \cite{Gallager_book, CoverThomas_book}, one has
$\MI(b; \r_1, \bpsi)=
\MI(b; z_1, \bpsi)$.
Therefore,  the coherent 
ergodic capacity
of \eqref{eq:r44} is given by
\be
\Cap_1 \eqdef  \sup_{f(b) \in \mathcal{I}_{b}}
\frac{\MI(b; z_1, \bpsi)}{M}
\quad \text{(in b/s/Hz)}
\label{eq:cap-1111111}
\ee
where $\mathcal{I}_{b}$ is the set of admissible input distributions
$f(b)$ fulfilling {\em both} the variance constraint $E(|b|^2)=\sigma_b^2$
and, according to \eqref{eq_gamma_symbol},
the amplitude constraint $|b| \le 1$ almost surely (a.s.).
We remember that, since the average of a
random variable cannot exceed its maximal value, the amplitude
constraint implies that $\sigma_b^2 \le 1$.

We observe that \eqref{eq:z1} is
a conditionally Gaussian channel, given $b$ and
$\bpsi$.
It was shown in \cite{Smith} that the capacity-achieving
input distribution for conditional Gaussian channels under
variance and amplitude constraints  is {\em discrete}
with a {\em finite} number of mass points.
Therefore,  there  is  no  loss  of  generality
in confining $f(b)$ to the set of discrete
distributions.
To this goal, let $b$ be a discrete random variable taking on the
value $\beta_q \in \mathcal{B}$ with probability
$p_q$,  for each $q \in \mathcal{Q}$, such that
$|\beta_q| \le 1$, $E(|b|^2)=\sigma_b^2$,
and $\sum_{q \in \mathcal{Q}} p_q =1$.
Using the same arguments of
Subsection~\ref{sec:analysis-legacy}, one gets
\be
\MI(b; z_1, \bpsi) =
\MI(b; z_1 \, | \, \bpsi)
=
\Es_{\bpsi}\left[\MI(b; z_1 \, | \, \bpsi= \bxi)\right]  \: .
\label{eq:MI-1}
\ee
For the discrete input $b$, the mutual information
$\MI(b; z_1 \, | \, \bpsi= \bxi)$
is given by
\be
\MI(b; z_1 \, | \, \bpsi= \bxi) =
\ent(z_1 \, | \, \bpsi= \bxi)-\ent(z_1 \, | \, b, \bpsi= \bxi)
\label{eq:II1}
\ee
where 
\be
\ent(z_1 \, | \, \bpsi= \bxi) =-\int_{\Cset} f_{z_1 \,|\, \bpsi=\bxi}(x) \, 
\log f_{z_1 \,|\, \bpsi=\bxi}(x) \, {\rm d}x
\ee
is the differential entropy \cite{Gallager_book, CoverThomas_book}
of  $z_1 \,|\, \bpsi=\bxi$, whereas 
\be
\ent(z_1 \, | \, b, \bpsi= \bxi)=\ent(\alpha \, \|\bpsi \|^2 \, b +
\bpsi^\herm \,{\v}_1 \, | \, b, \bpsi= \bxi) = 
\ent(\bpsi^\herm \,{\v}_1 \, | \, b, \bpsi= \bxi) = \ent(\bxi^\herm \,{\v}_1)
\ee
turns out to be the differential entropy of
$\bxi^\herm \, \v_1 \sim \CN(0, \sigma_{v_1}^2 \|\bxi\|^2)$,
which is given (see, e.g., \cite{Nee}) by $\ent(\bxi^\herm \,{\v}_1)=\log (\pi e \, \Es[|\bxi^\herm \, \v_1|^2])$.
It is noteworthy that, given $\bpsi=\bxi$, the output distribution 
\be
f_{z_1 \,|\, \bpsi=\bxi}(x) = \sum_{q=1}^{Q} p_q \, 
f_{z_1 \,|\, b=\beta_q, \bpsi=\bxi}(x) 
\label{eq:pdfz1}
\ee
is a Gaussian mixture since $z_{1} \, | \, b=\beta_q, \bpsi=\bxi \sim \CN(\alpha \, \|\bxi\|^2 \, \beta_q, \sigma_{v_1}^2 \|\bxi\|^2)$. By virtue of \eqref{eq:II1}, the optimization problem \eqref{eq:cap-1111111} is equivalent to the supremization of
$\Es_{\bpsi}[\ent(z_1 \, | \, \bpsi= \bxi)]$
under the variance and amplitude constraints.
However, the entropy $\ent(z_1 \, | \, \bpsi= \bxi)$
cannot be calculated in closed form due to the logarithm of
a sum of exponential functions.
As a consequence, an analytical expression for the optimizing
probability mass function (pmf) of $b$ is not  available for
the general case, neither there exists a  closed-form formula  for
the corresponding capacity.
Henceforth, upper and lower bounds on $\Cap_1$ given by \eqref{eq:cap-1111111}
are developed in the subsequent subsections.

\vspace{3mm}
\subsubsection{Upper bound on the capacity $\Cap_1$}
\label{sec:upper}

An upper bound on the ergodic capacity $\Cap_1$ can be
obtained by resorting to the maximum-entropy theorem for
complex random variables \cite{Nee}, which allows one to
state that
\be
\ent(z_1 \, | \, \bpsi= \bxi)  \le \log\left(\pi e \, \Es\left[|z_1|^2 \, \big | \, \bpsi= \bxi\right]\right) =
\log\left(\alpha^2  \, \sigma_b^2 \, \|\bxi\|^4+
\sigma_{v_1}^2 \|\bxi\|^2 \right) \: .
\label{eq:max-entropy}
\ee
By substituting  \eqref{eq:max-entropy} in \eqref{eq:II1} and
accounting for \eqref{eq:cap-1111111}--\eqref{eq:MI-1}, one gets
the upper bound
\be
\Cap_1  \le \Cap_{1,\text{upper}} \eqdef
\frac{1}{M} \,
\Es \left[ \log \left( 1 +  \SNRB  \, \Theta_{121} \right)\right]
\label{eq:C4-1}
\ee
with $\SNRB \eqdef \alpha^2 \sigma_b^2/\sigma_{v_1}^2$ and
\be
\Theta_{121} \eqdef \sum_{m=0}^{M-1} |s^{(m)}|^2 \, |\Psi_{12}(m)|^2
|\Psi_{21}(m)|^2 \:.
\label{eq:Theta}
\ee
It can be shown that,
as $Q$ grows,  $\Cap_1$ approaches $\Cap_{1,\text{upper}}$
exponentially fast \cite{Wu}.

In the general case, the evaluation of the
expectation in \eqref{eq:C4-1}
is significantly complicated and will be
numerically carried out in Section~\ref{sec:simul}.
Herein,  we shall
resort to a simpler asymptotic analysis by assuming that
$M$ is sufficiently large.
It follows from the law of large numbers \cite{Papoulis} that,
as $M$ gets large, the random variable $\Theta_{121}/M$ converges
a.s. to $\sigma_s^2 \, \sigma_{12}^2 \, \sigma_{21}^2$.
Hence, observing that,
according to the considered path-loss model, it results that
$\sigma_{12}^2=\sigma_{21}^2$ since
$d_{12}=d_{21}$,  in the large $M$ limit, one can write\sfootnote{For
any value of $M$, eq.~\eqref{eq:C4-2} is
as an upper bound on \eqref{eq:C4-1}:
by Jensen's inequality,
$\Es [ \log \left( 1 + \SNRB \, \Theta_{121} \right)] \le
\log [ 1 + \SNRB \, \Es(\Theta_{121}) ]$, with
$\Es(\Theta_{121})=M \, \sigma_s^2 \, \sigma_{12}^2 \, \sigma_{21}^2$.
}
\be
\Cap_1 \le \Cap_{1,\text{upper}} |_{\text{$M \gg1$}} \eqdef
\frac{1}{M} \,
\log \left( 1 + \SNRB \, M \,  \sigma_s^2 \, \sigma_{12}^4 \right) =
\frac{1}{M} \,
\log \left[ 1 + \SNRB
\, \frac{M \, \sigma_s^2}{(d_{12}^2)^{\eta}} \right] \: .
\label{eq:C4-2}
\ee

\vspace{3mm}
{\em Remark 7}:
When $M$ is sufficiently large, the upper bound \eqref{eq:C4-1}
is a monotonically increasing function of
$\SNRB$ and $1/d_{12}$. In other words,
significant high values of $\Cap_1$ are obtained
when the BTx reflects a large part of the
incident EM wave and/or the  BTx
is very close to the LTx.

\vspace{3mm}

\subsubsection{Lower bound on the capacity $\Cap_1$}
\label{sec:lower}

By resorting to random coding arguments (see, e.g., \cite{Wilson}),
it can be shown that the {\em cut-off rate}, which is defined as
follows
\be
R_1 \eqdef \max_{p_1, p_2, \ldots, p_Q} - \log \int_{\Cset}
\left[ \sum_{q=1}^Q p_q \sqrt{f_{z_1\,|\, b=\beta_q, \bpsi=\bxi}(x)} \right]^2
\!\!\! {\rm d}x
\label{eq:cutoff-1}
\ee
is a lower bound on $\MI(b; z_1 \, | \, \bpsi= \bxi)$ at any SNR.

By using the properties of the logarithmic function,
we observe that the objective function in \eqref{eq:cutoff-1} can be
explicated as reported at the top of this page in \eqref{eq:II11},
\begin{figure*}[!t]
\normalsize
\begin{multline}
\!\!\!\!\!\!\!\!\!\!\!\!\!\!\!
- \log \int_{\Cset}
\left[ \sum_{q=1}^Q p_q \sqrt{f_{z_1\,|\, b=\beta_q, \bpsi=\bxi}(x)} \right]^2
{\rm d}x  =
- \log \sum_{q_1=1}^Q \sum_{q_2=1}^Q p_{q_1} \, p_{q_2}
\int_{\Cset} \frac{1}{\pi \sigma_{v_1}^2 \|\bxi\|^2}
\, e^{-\frac{\left|x-\alpha \, \|\bxi\|^2 \, \beta_{q_1} \right|^2+
\left|x-\alpha \, \|\bxi\|^2 \, \beta_{q_2} \right|^2}
{2 \, \sigma_{v_1}^2 \|\bxi\|^2}}
\, {\rm d}x
\\  =
- \log \sum_{q_1=1}^Q \sum_{q_2=1}^Q p_{q_1} \, p_{q_2}
\, e^{-\frac{\alpha^2 \, \|\bxi\|^2 \,
\left|\beta_{q_1}-\beta_{q_2}\right|^2}{4 \, \sigma_{v_1}^2}}
\int_{\Cset} \frac{1}{\pi \sigma_{v_1}^2 \|\bxi\|^2} \,
e^{- \frac{\left|x-\alpha \, \|\bxi\|^2 \frac{\beta_{q_1}
+\beta_{q_2}}{2}\right|^2}{\sigma_{v_1}^2 \|\bxi\|^2}} \,
{\rm d}x
\\  =
- \log \sum_{q_1=1}^Q \sum_{q_2=1}^Q p_{q_1} \, p_{q_2}
\, e^{-\frac{\alpha^2 \, \|\bxi\|^2 \,
\left|\beta_{q_1}-\beta_{q_2}\right|^2}{4 \, \sigma_{v_1}^2}}
\label{eq:II11}
\end{multline}
\hrulefill
\end{figure*}
where
the last but one equality is obtained by completion of
the square in the exponent, whereas the last integral is $1$ for any choice of
the symbol set $\mathcal{B}$, since it is recognized as the
integral of a univariate complex Gaussian pdf. Eq.~\eqref{eq:II11} is valid for
{\em any} finite-size symbol constellation, such as
quadrature-amplitude modulation (QAM),
PSK, orthogonal, lattice-type, or other.
It is verified \cite{Wilson} that, for symbol constellations
where the set of distances to other neighbors is
invariant to the choice of the reference point, e.g.,
PSK and orthogonal modulations, the equiprobable
assignment on the backscatter symbols (i.e., $p_q=1/Q
\,\, \forall q \in \mathcal{Q}$) maximizes
\eqref{eq:II11}. Therefore, remembering
\eqref{eq:cap-1111111}, \eqref{eq:MI-1},  and \eqref{eq:cutoff-1},
one yields $\Cap_1 \ge \Cap_{1,\text{lower}}$, with
\be
\Cap_{1,\text{lower}} \eqdef
\frac{\log Q - \Es \left[
\log \left(1 + \sum_{q=2}^Q e^{- \Theta_{121} \, \SNRB \,
\frac{\left|\beta_{1}-\beta_{q}\right|^2}{4 \, \sigma_b^2}} \right)
\right]}{M}
\label{eq:lowbouc1}
\ee
where $\SNRB$ and $\Theta_{121}$ have been defined
in Subsection~\ref{sec:upper}.
The further lower bound
\be
\Cap_{1,\text{lower}} \ge
\frac{\log Q - \Es \left[
\log \left(1 + (Q-1) \, e^{- \Theta_{121} \, \SNRB \,
\frac{\deltamin^2}{4 \, \sigma_b^2}} \right)
\right]}{M}
\label{eq:lowbouc1f}
\ee
can be obtained by noting that
$|\beta_{1}-\beta_{q}|^2 \ge \deltamin^2$
for each $q \in \mathcal{Q}$,
where $\deltamin \eqdef \min_{q_1 \neq q_2 \in \mathcal{Q}} |\beta_{q_1}-\beta_{q_2}|$ is the minimum distance between any two data symbols in the signal constellation $\mathcal{B}$.
By invoking again the law of large numbers \cite{Papoulis},
in the large $M$ limit, the following asymptotic expressions
of $\Cap_{1,\text{lower}}$ and its lower bound
\eqref{eq:lowbouc1f} hold
\be
\Cap_{1,\text{lower} |_{\text{$M \gg1$}}}  \eqdef
\frac{\log Q -
\log \left(1 + \sum_{q=2}^Q
e^{- \frac{\sigma_s^2 \, M \, \SNRB \,
\left|\beta_{1}-\beta_{q}\right|^2}{4 \, \sigma_b^2 \, (d_{12}^2)^{\eta}}} \right)}{M}    \ge
\frac{\log Q  -
\log \left(1 + (Q-1) \,
e^{- \frac{\sigma_s^2 \, M \, \SNRB \, \deltamin^2}{4 \, \sigma_b^2 \, (d_{12}^2)^{\eta} }} \right)}{M}  \: .
\ee

\vspace{3mm}
{\em Remark 8}:
The lower bound $\Cap_{1,\text{lower}}$ approaches
$(\log Q)/M$ as $\SNRB$ increases or the distance $d_{12}$
between the LTx and the BTx decreases. On the other hand,
when $x \to 0$,  the function $\log(1+ A \, e^{- B x})$
can be approximated using the first two terms of its Mac Laurin
series expansion, i.e.,
$\log(1+ A \, e^{- B \, x}) \approx \log(1+A) - A \, B \, (1+A)^{-1} x$,
in the  low-SNR regime $\SNRB \to 0$ or when $d_{12} \to + \infty$,
hence getting
\be
\Cap_{1,\text{lower} |_{\text{$M \gg1$}}}  \to
\left(1-\frac{1}{Q}\right) \,
\frac{\sigma_s^2 \, \SNRB \, \deltamin^2}{4 \, \sigma_b^2 \, (d_{12}^2)^{\eta} }
\ee
that is, the capacity increases linearly
with $\SNRB$ and monotonically decreases as
the distance $d_{12}$ raises.

\subsection{The BRx and LTx are spatially-separated nodes}
\label{sec:distinct}

We consider the scenario where the
LTx and BRx are spatially-separated nodes, which is
the situation considered in \cite{Liu}. In this
case, taking into account the aforementioned
simplifying assumptions
\eqref{eq:lbb} and $\Lcp=L_b$,
the reference signal model \eqref{eq:r4}--\eqref{eq:d4}
becomes
\be
\R_{b} \, \tilde{\r}_4(n)  =
\alpha \left[ e^{j \frac{2 \pi}{M} \nu n P} \, \R_b \, \bSigma_{\nu} \, \tilde{\C}_{24}^{(0)} \, \tilde{\C}_{12}^{(0)} \,  \u(n) \right] b(n) + \d_4(n)
\label{eq:r444}
\ee
with
\be
\d_4(n) =
e^{j \frac{2 \pi}{M} \nu n P} \, \R_b \, \bSigma_{\nu} \,
\tilde{\C}_{14}^{(0)} \, \u(n) + \R_b \, \tilde{\v}_4(n) \in \Cset^{M}
\label{eq:d444} \: .
\ee
Compared to the case studied in
Subsection~\ref{sec:coincide},
there are two key differences: (i) the receiver has no
knowledge of the data block $\u(n)=\Tcp \, \Widft \, \s(n)$ transmitted by the LTx;
(ii) there is a nonzero CFO $\nu$ between the received carrier and
the local sinusoids used for signal demodulation.

If the BRx does not have any {\em a priori} knowledge regarding the
legacy transmission, recovery of $b(n)$ can be accomplished at the BRx by resorting to {\em noncoherent} detection rules.
The noncoherent ergodic capacity of
\eqref{eq:r444}--\eqref{eq:d444} is given by the supremum of
the mutual information $\MI[b; \R_{b} \, \tilde{\r}_4(n)]$ over
the set $\mathcal{I}_b$ of admissible input distribution
satisfying both the variance and amplitude constraints.
Evaluation of the noncoherent ergodic capacity with
only a variance constraint has been
studied in \cite{Marzetta1999,Hock2000, Zheng2002}
under the assumption that the channel matrix
[corresponding to $e^{j \frac{2 \pi}{M} \nu n P} \, \R_b \, \bSigma_{\nu} \, \tilde{\C}_{24}^{(0)} \, \tilde{\C}_{12}^{(0)} \,  \u(n)$ in our framework]
and noise [corresponding to $\d_4(n)$ in our framework] follow
a Gaussian distribution. In the case under study, evaluation of
the noncoherent ergodic capacity is further complicated by
the non-Gaussian nature of both $e^{j \frac{2 \pi}{M} \nu n P} \, \R_b \, \bSigma_{\nu} \, \tilde{\C}_{24}^{(0)} \, \tilde{\C}_{12}^{(0)} \,  \u(n)$
and $\d_4(n)$, as well as by the amplitude constraint $|b| \le 1$.

To avoid incurring the data-rate penalty of the noncoherent communication scheme,
we study the case where, besides having knowledge of the training symbols transmitted
by the BTx, the BRx additionally knows the pilot symbols
sent by the LTx in each frame. Under this assumption, following the same
protocol outlined in Subsubsection~\ref{sec:coincide}, during the silent
period of the BTx (i.e., when $\alpha=0$),
the BRx receives the signal $\d_4(n)$, from which it can estimate
the CFO $\nu$ and the parameters of the channel matrix
$\tilde{\C}_{14}^{(0)}$ by resorting to standard estimators \cite{Morelli, Mengali}.
However, it should be observed that the interference contribution
$e^{j \frac{2 \pi}{M} \nu n P} \, \R_b \, \bSigma_{\nu} \,
\tilde{\C}_{14}^{(0)} \, \u(n)$ cannot be subtracted
from \eqref{eq:r444} since the information-bearing data in $\u(n)$
are unknown at the BRx (only the pilots and their locations are
assumed to be known). Once $\nu$ has been estimated,
the vector $\tilde{\r}_4(n)$ can be counter-rotated
at the angular  speed  $2 \pi \nu/M$, thus yielding
\be
\r_4 \eqdef \R_{b} \, \tilde{\r}_4  =
\alpha \left( \Widft \, \bCcal_{12} \, \bCcal_{24} \,  \s \right) b
+ \Widft \, \bCcal_{14} \, \s + \v_4
\label{eq:r4444}
\ee
with $\v_4 \eqdef \R_{b} \, \tilde{\v}_4 \in \Cset^M$,
where $\s  \sim \CN(\bm{0}_{M}, \Ps \, \I_{M})$
is the capacity-achieving distribution for the legacy system
(see Section~\ref{sec:analysis-legacy}) and we have again omitted the dependence of the frame index $n$.
Since $\bOmega_{124} \eqdef \bCcal_{12} \, \bCcal_{24}
\in \Cset^{M \times M}$ and
$\bOmega_{14} \eqdef \bCcal_{14} \in \Cset^{M \times M}$ are known
but $\s$ is unknown, we refer to \eqref{eq:r4444} as the
{\em partially-coherent} channel model.
The partially-coherent ergodic capacity
of \eqref{eq:r4444} is given by
\be
\Cap_4 \eqdef  \sup_{f(b) \in \mathcal{I}_{b}}
\MI(b; \r_4, \bOmega_{124}, \bOmega_{14})
\label{eq:Cap4-10}
\ee
where $\mathcal{I}_{b}$ is the set of admissible input distributions
fulfilling $E(|b|^2)=\sigma_b^2$ and $|b| \le 1$ a.s., and
\be
\MI(b; \r_4, \bOmega_{124}, \bOmega_{14})  =
\MI(b; \r_4 \, | \, \bOmega_{124}, \bOmega_{14})
=
\Es_{\bOmega_{124}, \bOmega_{14}}\left[\MI(b; \r_4 \, | \,
\bOmega_{124}=\bXi_{124}, \bOmega_{14}=\bXi_{14}) \right] \:.
\label{eq:Cap4-20}
\ee
Similarly to the case studied in Subsection~\ref{sec:coincide},
closed-form expressions for $\Cap_4$
and the corresponding capacity-achieving discrete distribution $f(b)$
are unavailable. Therefore, we derive upper and lower
bounds on $\Cap_4$.

\vspace{3mm}
\subsubsection{Upper bound on the capacity $\Cap_4$}
\label{sec:upper-4}

An upper bound on $\Cap_4$ can be obtained by assuming
that the BRx has the additional perfect knowledge of
$\s$. Indeed, by using the chain rule for mutual information
\cite{Gallager_book, CoverThomas_book}, it can be
proven that
\be
\MI(b; \r_4 \, | \, \bOmega_{124}, \bOmega_{14}, \s) =
\MI(b; \r_4, \, | \, \bOmega_{124}, \bOmega_{14}) +
\MI(b; \s \, | \, \r_4, \bOmega_{124}, \bOmega_{14}) \ge
\MI(b; \r_4, \, | \, \bOmega_{124}, \bOmega_{14})
\label{eq:dissss}
\ee
since $\MI(b; \r_4, \, | \, \bOmega_{124}, \bOmega_{14}) \ge 0$
by definition. Moreover, because subtracting a constant does
not change mutual information \cite{Gallager_book, CoverThomas_book},
one has
\barr
\MI(b; \r_4 \, | \, \bOmega_{124}, \bOmega_{14}, \s)  & =
\MI(b; \r_4 - \Widft \, \bOmega_{14}\, \s \, | \, \bOmega_{124}, \bOmega_{14}, \s)
\nonumber \\ &
= \Es_{\bOmega_{124}, \bOmega_{14}, \s}
\left[\MI(b; \r_4 - \Widft \, \bOmega_{14}\, \s \, | \, \bOmega_{124}=\bXi_{124},
\bOmega_{14}=\bXi_{14}, \s=\a) \right]
\label{eq:IIII}
\earr
that is, since the BRx knows $\bOmega_{14}$ and $\s$, it can estimate
$b$ by subtracting $\Widft \, \bOmega_{14}\, \s$
from \eqref{eq:r4444}, hence yielding
$\hat{\r}_{4}  \eqdef \r_4 - \Widft \, \bOmega_{14}\, \s=  \alpha \, \Widft \, \bOmega_{124} \,  \s \, b + \v_4$.
It follows that
\begin{multline}
\MI(b; \r_4 - \bOmega_{14}\, \s \, | \, \bOmega_{124}=\bXi_{124},
\bOmega_{14}=\bXi_{14}, \s=\a)  =
\ent(\hat{\r}_{4} \, | \, \bOmega_{124}=\bXi_{124},
\bOmega_{14}=\bXi_{14}, \s=\a)
\\ -
\ent(\hat{\r}_{4} \, | \, b, \bOmega_{124}=\bXi_{124},
\bOmega_{14}=\bXi_{14}, \s=\a)
\label{eq:MIChi}
\end{multline}
where $\ent(\hat{\r}_{4} \, | \, b, \bOmega_{124}=\bXi_{124},
\bOmega_{14}=\bXi_{14}, \s=\a)=\ent(\v_4)= M \log(\pi e \, \sigma_{v_4}^2)$.
As a consequence of the maximum-entropy theorem for
complex random variables \cite{Nee}, one
can obtain a further upper bound on
\eqref{eq:MIChi} by observing that
\barr
\ent(\hat{\r}_{4} \, | \, \bOmega_{124}=\bXi_{124},
\bOmega_{14}=\bXi_{14}, \s=\a) & \le \log \left\{(\pi e)^M \det \left( \Es\left[\hat{\r}_{4} \, \hat{\r}_{4}^\herm \, \big| \, \bOmega_{124}=\bXi_{124},
\bOmega_{14}=\bXi_{14}, \s=\a \right]\right) \right\}
\nonumber \\ & =
\log \left[ (\pi e \, \sigma_{v_4}^2)^M  \left( 1+ \SNRBB \,
\|\bXi_{124} \,  \a\|^2\right) \right]
\label{eq:ent4}
\earr
with $\SNRBB \eqdef \alpha^2 \sigma_b^2/\sigma_{v_4}^2$,
where we have used the facts \cite{Horn} that: (i) $\det(\A \, \B)=\det(\A) \, \det(\B)$ for arbitrary nonsingular matrices $\A \in \Cset^{n \times n}$ and $\B \in \Cset^{n \times n}$; (ii) $\det(\Widft) \, \det(\Wdft)=1$; (iii)
for arbitrary vectors
$\x \in \Cset^n$ and $\y \in \Cset^n$, $\det(\I_n + \x \, \y^\herm)=1+\x^\herm \y$.
Henceforth, accounting for \eqref{eq:dissss}--\eqref{eq:ent4}, it results
from \eqref{eq:Cap4-10} that
\be
\Cap_4  \le \Cap_{4,\text{upper}} \eqdef
\frac{1}{M} \,
\Es \left[ \log \left( 1 +  \SNRBB  \, \Theta_{124} \right)\right]
\label{eq:C4-11}
\ee
with
\be
\Theta_{124} \eqdef \sum_{m=0}^{M-1} |s^{(m)}|^2 \, |\Psi_{12}(m)|^2 |\Psi_{24}(m)|^2 \:.
\label{eq:Thetaa}
\ee
Such an upper bound is achieved when the BRx is
able to reliably estimate the legacy symbols and $Q \to +\infty$.
It should be noted that \eqref{eq:C4-11} is similar to \eqref{eq:C4-1}.
Thus, the asymptotic analysis reported in Subsection~\ref{sec:upper} soon after
\eqref{eq:C4-1} can be applied to \eqref{eq:C4-11}
with minor modifications. In particular, in the large $M$ limit,
one obtains
\be
\Cap_4 \le \Cap_{4,\text{upper}} |_{\text{$M \gg1$}} \eqdef
\frac{1}{M} \,
\log \left( 1 + \SNRBB \, M \,  \sigma_s^2 \, \sigma_{12}^2 \, \sigma_{24}^2 \right) =
\frac{1}{M} \,
\log \left[ 1 + \SNRBB
\, \frac{M \, \sigma_s^2}{(d_{12} \, d_{24})^{\eta}} \right]
\label{eq:C44-2}
\ee
where,  by virtue of the Carnot's cosine law, the distances $d_{12}$
and $d_{24}$ are related by
$d_{24} = (d_{12}^2+d_{14}^2 - 2 \, d_{12} \, d_{14} \, \cos \theta)^{1/2}$,
with $\theta$ being the angle opposite to the $2 \to 4$ link
(see Fig.~\ref{fig:figure_1}).

\vspace{3mm}
{\em Remark 9}: For a fixed value of $d_{14}$ and $\theta$,
the capacity
$\Cap_{4,\text{upper}} |_{\text{$M \gg1$}}$  as a function of
$d_{12}$ exhibits the same behavior of $\Delta \Cap_3$
(see Remark~6). In a nutshell, when $\theta \in \mathcal{A}$,
the upper bound $\Cap_{4,\text{upper}} |_{\text{$M \gg1$}}$ is a strictly decreasing function of $d_{12}/d_{14}$, whereas, when
$\theta \not \in \mathcal{A}$,
it monotonically increases for
$d_{\text{min}}(\theta) \le {d_{12}}/{d_{14}} \le d_{\text{max}}(\theta)$,
otherwise, it monotonically decreases. In other words, if the BRx reliably
estimates the legacy symbols, in the former case,
the capacity $\Cap_4$ of the backscatter system decreases while
the BTx is departing from the LTx, whereas, in latter one, it increases  as the BTx
approaches to either the LTx or an intermediate point between the LTx
and the BRx.

\vspace{3mm}
\subsubsection{Lower bound on the capacity $\Cap_4$}
\label{sec:lower-4}

As in Subsection~\ref{sec:lower}, we rely on the fact that
\be
R_4 \le \MI(b; \r_4 \, | \,
\bOmega_{124}=\bXi_{124}, \bOmega_{14}=\bXi_{14})
\label{eq:R4}
\ee
where $R_4$ is the cut-off rate when the
backscatter symbols are assumed to be equiprobale, that is,
\be
R_4 \eqdef - \log \int_{\Cset^M}
\left[ \frac{1}{Q} \sum_{q=1}^Q \sqrt{f_{\r_4 \,|\, b=\beta_q, \bOmega_{124}=\bXi_{124}, \bOmega_{14}=\bXi_{14}}(\x)} \right]^2
\!\!\! {\rm d}\x \: .
\label{eq:cutoff-4}
\ee
Eq.~\eqref{eq:r4444} shows that
$\r_4 \,|\, b=\beta_q, \bOmega_{124}=\bXi_{124}, \bOmega_{14}=\bXi_{14} \sim \CN[\mathbf{0}_M, \K_{4}(\beta_q,\bXi_{124}, \bXi_{14})]$,
with
$\K_{4}(\beta_q,\bXi_{124}, \bXi_{14})  \eqdef
\Es(\r_{4} \, \r_{4}^\herm \,\big|\, b=\beta_q, \bOmega_{124}=\bXi_{124}, \bOmega_{14}=\bXi_{14}) = \Widft \, \R_{4}(\beta_q,\bXi_{124}, \bXi_{14}) \, \Wdft $, where
$\R_{4}(\beta_q,\bXi_{124}, \bXi_{14})   \eqdef
\alpha^2 \, \sigma_s^2 \,
\bXi_{124} \, \bXi_{124}^*  \, |\beta_q|^2  +
\alpha\, \sigma_s^2 \, \bXi_{124} \, \bXi_{14}^* \, \beta_q
+ \alpha\, \sigma_s^2 \, \bXi_{124}^*  \, \bXi_{14} \, \beta_q^*
+ \sigma_s^2 \, \bXi_{14} \, \bXi_{14}^* + \sigma_{v_4}^2 \, \I_M$
is a diagonal matrix.
By using the properties of the determinant \cite{Horn},
we observe that $R_4$ can be
explicated as reported at the top of this page in \eqref{eq:II111},
\begin{figure*}[!t]
\normalsize
\begin{multline}
R_4 =
- \log \sum_{q_1=1}^Q \sum_{q_2=1}^Q  \frac{1}{Q^2}
\int_{\Cset^M} \frac{e^{- \frac{\x^\herm \left[\K^{-1}_{4}(\beta_{q_1}) + \K^{-1}_{4}(\beta_{q_2})\right]  \x}{2}}}{\pi^M
\sqrt{\det\left[\K_{4}(\beta_{q_1})\right]
\det \left[ \K_{4}(\beta_{q_2}) \right ]}}
\, {\rm d}\x
\\  =
- \log \sum_{q_1=1}^Q \sum_{q_2=1}^Q
\frac{ \det\left\{ 2 \,
\left[\K^{-1}_{4}(\beta_{q_1}) 
+ \K^{-1}_{4}(\beta_{q_2})\right]^{-1}
\right\}}{Q^2 \sqrt{\det\left[\K_{4}(\beta_{q_1})\right]
\det \left[
\K_{4}(\beta_{q_2}) \right ]}} \,
\int_{\Cset^M} \frac{e^{- \frac{\x^\herm \left[\K^{-1}_{4}(\beta_{q_1}) + \K^{-1}_{4}(\beta_{q_2})\right]  \x}{2}}}{\pi^M \, \det\left\{ 2 \,
\left[\K^{-1}_{4}(\beta_{q_1}) + \K^{-1}_{4}(\beta_{q_2})\right]^{-1}
\right\}} \,
\, {\rm d}\x
\\  =
\log Q - \log \left[ 1+ \frac{2^M}{Q} \sum_{q_1=1}^Q
\sum_{\shortstack{\footnotesize $q_2=1$ \\
\footnotesize $q_2 \neq q_1$}}^Q
\frac{1}{\sqrt{\det\left[\R_{4}(\beta_{q_1}) \right] \det \left[ \R_{4}(\beta_{q_2}) \right ]}
\det[\R^{-1}_{4}(\beta_{q_1}) + \R^{-1}_{4}(\beta_{q_2})]
}
\right] \\
\label{eq:II111}
\end{multline}
\hrulefill
\end{figure*}
where we have omitted to explicitly indicate 
the dependence of $\K_4(\cdot)$ and $\R_4(\cdot)$
on $\bXi_{124}$ and  $\bXi_{14}$, and 
the last integral is the
hypervolume of a multivariate complex Gaussian pdf.
By virtue of \eqref{eq:Cap4-20} and \eqref{eq:R4},
the capacity \eqref{eq:Cap4-10} is lower bounded as
shown at the top of this page in \eqref{eq:C4-11-low},
\begin{figure*}[!t]
\normalsize
\be
\Cap_4  \ge \Cap_{4,\text{lower}} \eqdef
\frac{\log Q - \Es \left[ \log \left(1+ \frac{2^M}{Q}
\displaystyle \sum_{q_1=1}^Q
\sum_{\shortstack{\footnotesize $q_2=1$ \\
\footnotesize $q_2 \neq q_1$}}^Q \prod_{m=0}^{M-1}
\frac{\sqrt{\Lambda_{q_1}(m) \,
\Lambda_{q_2}(m)}}{
\Lambda_{q_1}(m) + \Lambda_{q_2}(m)}\right) \right]}{M}
\label{eq:C4-11-low}
\ee
\hrulefill
\end{figure*}
with
\be
\Lambda_{q}(m) \eqdef
\alpha^2 \, \sigma_s^2 \, |\Psi_{12}(m)|^2 \, |\Psi_{24}(m)|^2 \,
 |\beta_q|^2
+ 2 \, \alpha\, \sigma_s^2 \, \Re \left \{\Psi_{12}(m) \, \Psi_{24}(m) \, \Psi_{14}^*(m) \, \beta_q \right \}
+ \sigma_s^2 \, |\Psi_{14}(m)|^2 + \sigma^2_{v_4} \: .
\label{eq:lambdaqm}
\ee
In addition to noise, another additive source of
performance degradation is the
interference generated
by the legacy system over the $1 \to 4$ link,
which may seriously limit the achievable rates
of the backscatter system in the high-SNR region.

\vspace{3mm}
{\em Remark 10}: It is verified from
\eqref{eq:C4-11-low} that $\Cap_{4,\text{lower}} \to 0$
if $\Lambda_{q_1}(m) \to \Lambda_{q_2}(m)$
for each $q_1 \neq q_2 \in \mathcal{Q}$.
For instance, this happens when the second and third summands in
the RHS of \eqref{eq:lambdaqm} are dominant over the
first and second ones, i.e., when  interference and/or noise dominates
the backscatter signal.

\vspace{3mm}
The dependence of the
$\Cap_{4,\text{lower}}$ on the distance $d_{12}$ between
the LTx and the BTx is not easily deduced from \eqref{eq:C4-11-low}
and such a behavior will be studied numerically in Section~\ref{sec:simul}.
To gain some useful insights, we consider the special case of a
$2$-PSK (i.e., BPSK), where $\beta_1=-\beta_2=1$.
In this case, eq.~\eqref{eq:C4-11-low} becomes
\barr
\Cap_{4,\text{lower}}^{\text{bpsk}} & =\frac{1}{M}- \frac{1}{M} \, \Es \left\{\log\left[1+ \prod_{m=0}^{M-1} \sqrt{1-\frac{\Lambda_1^2(m)}{\Lambda_2^2(m)}}\right] \right\} \ge
\frac{1}{M}- \frac{1}{M} \, \log\left\{1+ \sqrt{\Es \left[\prod_{m=0}^{M-1} \left({1-\frac{\Lambda_1^2(m)}{\Lambda_2^2(m)}}\right) \right]} \right\}
\nonumber \\ & \approx
\frac{1}{M}- \frac{1}{M} \, \log\left\{1+ \sqrt{\prod_{m=0}^{M-1}
\left(
1-\Es \left[ \frac{\Lambda_1^2(m)}{\Lambda_2^2(m)}\right]\right)} \right\}
\label{eq:bpsk}
\earr
with
\barr
\Lambda_1(m) & \eqdef
\alpha^2 \, \sigma_s^2 \, |\Psi_{12}(m)|^2 \, |\Psi_{24}(m)|^2
+ \sigma_s^2 \, |\Psi_{14}(m)|^2 + \sigma^2_{v_4}
\\
\Lambda_2(m) & \eqdef
2 \, \alpha\, \sigma_s^2 \, \Re \left \{\Psi_{12}(m) \, \Psi_{24}(m) \, \Psi_{14}^*(m)  \right \}
\earr
where the inequality in \eqref{eq:bpsk} comes from the application
of the Jensens's inequality to the concave function $\log(1+\sqrt{x})$, whereas
the approximation is obtained by neglecting the correlation between
the random variables ${\Lambda_1^2(m_1)}/{\Lambda_2^2(m_1)}$
and ${\Lambda_1^2(m_2)}/{\Lambda_2^2(m_2)}$, for
$m_1 \neq m_2 \in \mathcal{M}$.
The first-order Taylor expansion of  $\Es
\left[{\Lambda_1^2(m)}/{\Lambda_2^2(m)}\right]$ leads to the further approximation\sfootnote{Let
$f(X,Y) \eqdef X/Y$ be a transformation of the two random variables $X$
and $Y$. Let $\mu_X \eqdef \Es(X)$ and $\mu_Y \eqdef \Es(Y)$,
the first-order Taylor approximation for $\Es[f(X,Y)]$ is given
by $\Es[f(X,Y)]=f(\mu_X,\mu_Y) +\Es [f_x'(\mu_X,\mu_Y) \, (X-\mu_X)]
+ \Es [f_y'(\mu_X,\mu_Y) \, (Y-\mu_Y)] = f(\mu_X,\mu_Y) =
\mu_X/\mu_Y$, where $f'_x(\cdot)$ and $f'_y(\cdot)$ are the partial derivatives of  the function $f(x,y)$ with respect to the real-valued variables $x$ and $y$, respectively.
}
$\Es
\left[{\Lambda_1^2(m)}/{\Lambda_2^2(m)}\right] \approx
{\Es[\Lambda_1^2(m)]}/{\Es[\Lambda_2^2(m)]}$, where in the
low-noise regime $\sigma^2_{v_4}/\sigma_s^2 \to 0$, one has
\be
J(d_{12}) \eqdef \lim_{\sigma^2_{v_4}/\sigma_s^2 \to 0} \frac{\Es[\Lambda_1^2(m)]}{\Es[\Lambda_2^2(m)]} =
\frac{1}{1+ \frac{2}{D(d_{12})}+ D(d_{12})}
\ee
where
$D(d_{12})  \eqdef \left({d_{12} \, d_{24}}/{d_{14}}\right)^\eta/\alpha^2$ is independent  of $m$, with
$d_{24} = (d_{12}^2+d_{14}^2 - 2 \, d_{12} \, d_{14} \, \cos \theta)^{1/2}$.
Therefore, when $\sigma^2_{v_4}/\sigma_s^2 \to 0$,
it results from \eqref{eq:bpsk} that\sfootnote{Using similar bounding/approximation techniques, a
lower bound on $\Cap_{4,\text{lower}}$ can be obtained for an arbitrary
$M$-ary backscatter signal constellation, which however
does not lend itself to easily interpretable results.}
\be
\Cap_{4,\text{lower}}^{\text{bpsk}} \gtrsim
\frac{1}{M}- \frac{1}{M} \, \log\left\{1+
\left[1- J(d_{12})\right]^{M/2} \right\} \: .
\ee

\vspace{3mm}
{\em Remark 11}:
For a fixed value of $d_{14}$ and $\theta$ (see Fig.~\ref{fig:figure_1}),
by using standard concepts of
mathematical analysis,\sfootnote{Details are omitted
in  the  interest  of
saving  space.} it can be shown that, if
$\theta \in \mathcal{A}$, then
$J(d_{12})$ is a {\em unimodal} function,
exhibiting a maximum when
$D(d_{12}) = \sqrt{2}$.
On the other hand,
when $\theta \not \in \mathcal{A}$, the function
$J(d_{12})$ is {\em multimodal} having multiple
local extrema points.

\section{Numerical performance analysis}
\label{sec:simul}

We present the Monte Carlo numerical analysis of the considered ambient
backscatter network to validate
and complete our theoretical analysis, with reference
to both legacy and backscatter systems.
All the ensemble averages (with respect to all the relevant fading
channels and information-bearing symbols) and the outage probability
of the legacy system
are evaluated through $10^6$ independent Monte Carlo runs.

In all the experiments, we adopted the following simulation setting.
With reference to the Cartesian plane in Fig.~\ref{fig:figure_1},
all the distances are normalized with respect to $d_{13}=1$.
Specifically,  the nodes $1$ (LTx) and $3$ (LRx) have coordinates equal to $(-0.5,0)$ and $(0.5,0)$,  respectively.
In all the plots where the distance $d_{12}$ varies,  
the node $2$ (BTx) moves along the line 
joining the nodes $1$ and $2$ (see Fig.~\ref{fig:figure_1}).
The multicarrier legacy system employs $M=32$ subcarriers and a CP of length $\Lcp=8$. The legacy symbols are
generated according to the corresponding capacity-achieving distribution
$\s \sim \CN(\bm{0}_{M}, \Ps \, \I_{M})$, with $\Ps=1$.
On the other hand, the symbols transmitted by the backscatter device
are equiprobably drawn from
BPSK, $4$-PSK (i.e., QPSK), and quaternary
amplitude-shift keying (ASK) signal
constellations, with average energy
$\sigma_b^2=1$.
The order of the discrete-time channels 
between the nodes is set equal to $L_{13} =L_{12}=L_{23}=3$, 
whereas the corresponding time offsets are fixed to 
$\theta_{13}=\theta_{12}=\theta_{23}=1$, respectively. Moreover, the path-loss exponent is chosen equal to $\eta=3$. For the evaluation of the outage probability of the legacy system, we chose $R_s=6$ b/s/Hz
in \eqref{eq:pout3}.

\begin{figure*}[t!]
\centering
\begin{minipage}[t]{0.45\textwidth}
\includegraphics[width=\linewidth, trim=20 20 40 20]{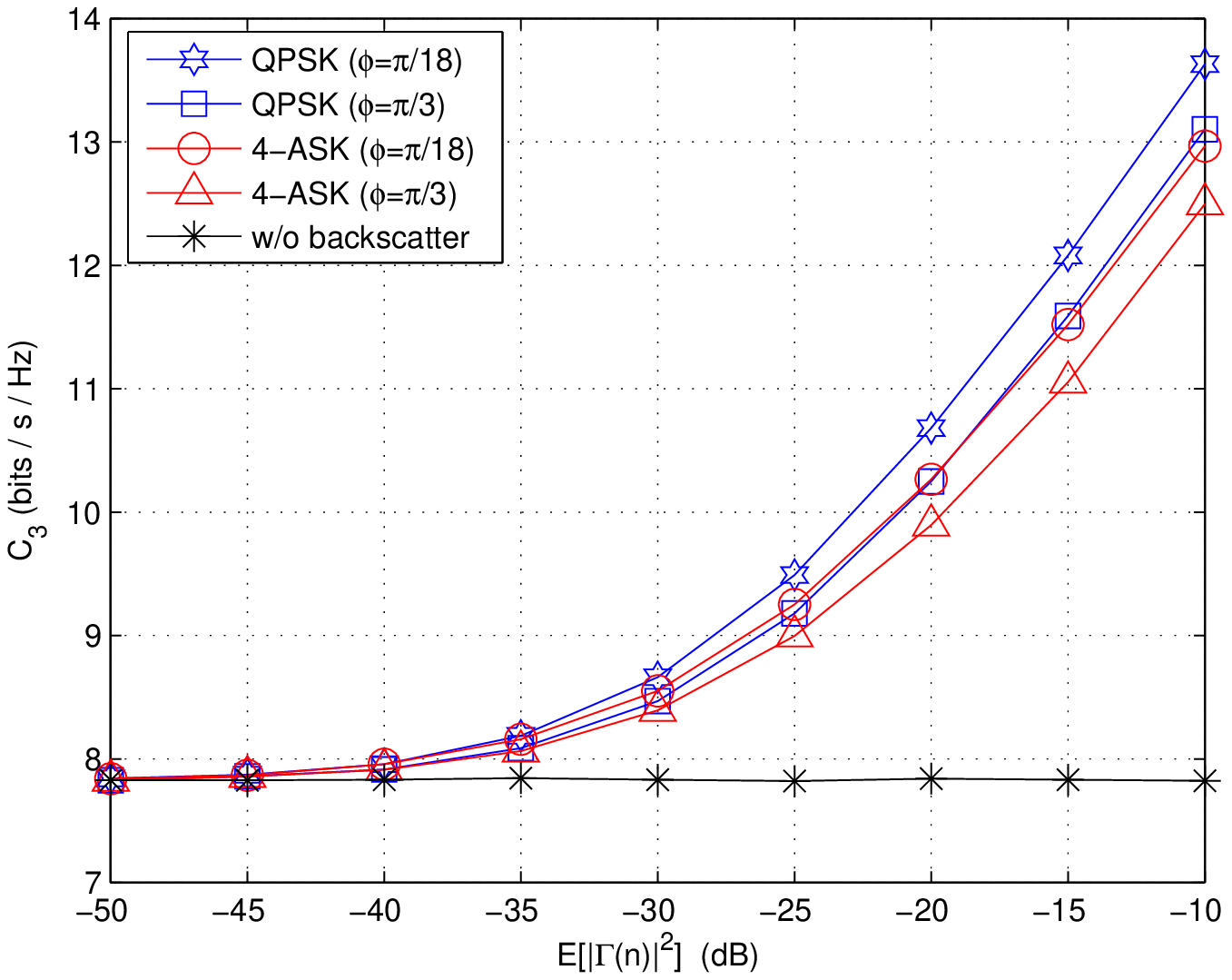}
\caption{Ergodic capacity of the legacy system versus
$\Es[\left|\Gamma(n)\right|^2]=\alpha^2$ for two backscatter signal constellations
and two values of the angle $\phi$.}
\label{fig:fig_3}
\end{minipage}%
\hspace{0.04\textwidth}%
\begin{minipage}[t]{0.45\textwidth}
\includegraphics[width=\linewidth, trim=20 20 40 20]{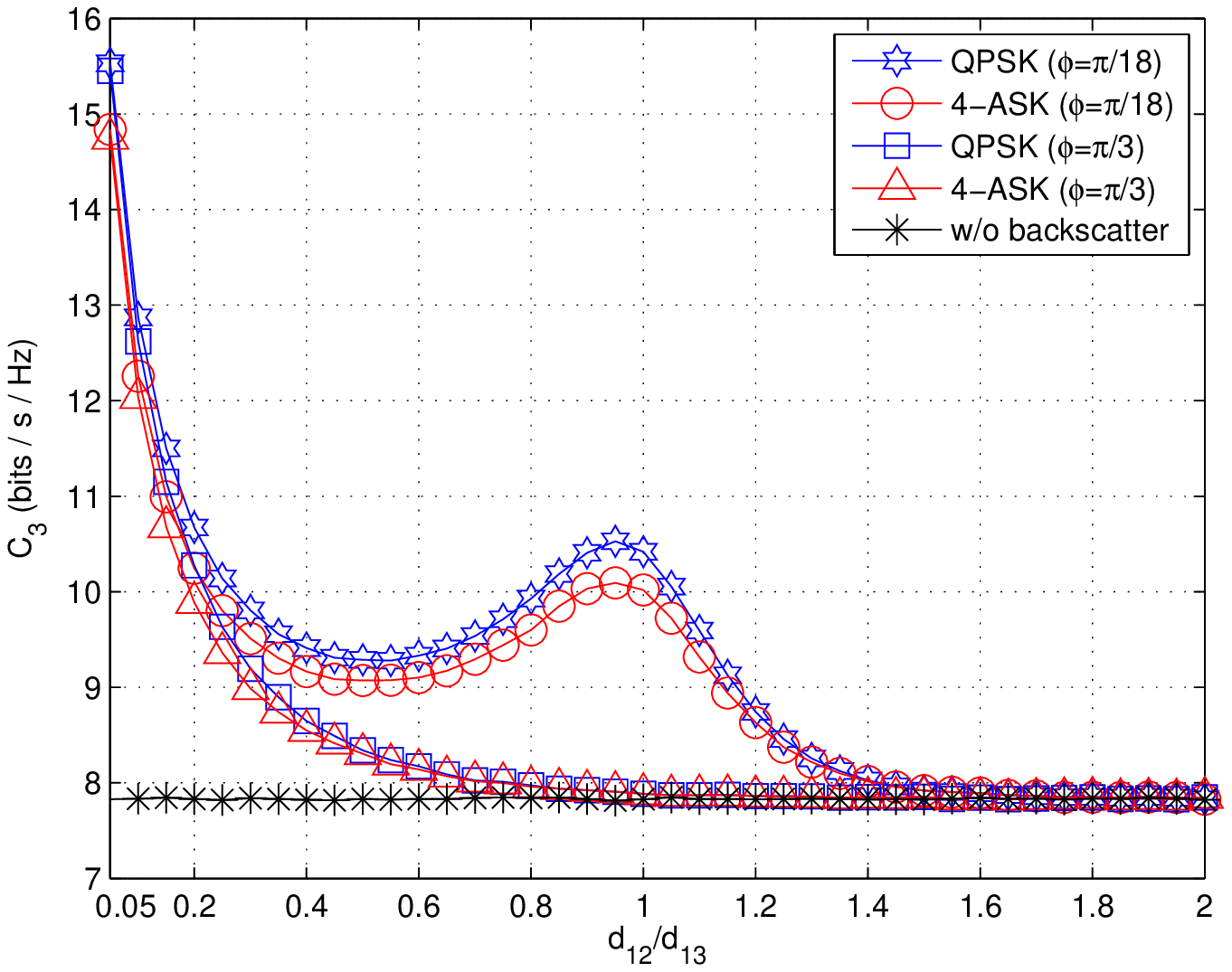}
\caption{Ergodic capacity of the legacy system versus $d_{12}/d_{13}$ for two backscatter signal constellations
and two values of the angle $\phi$.}
\label{fig:fig_4}
\end{minipage}
\end{figure*}
\begin{figure*}[t!]
\centering
\begin{minipage}[t]{0.45\textwidth}
\includegraphics[width=\linewidth, trim=20 20 40 20]{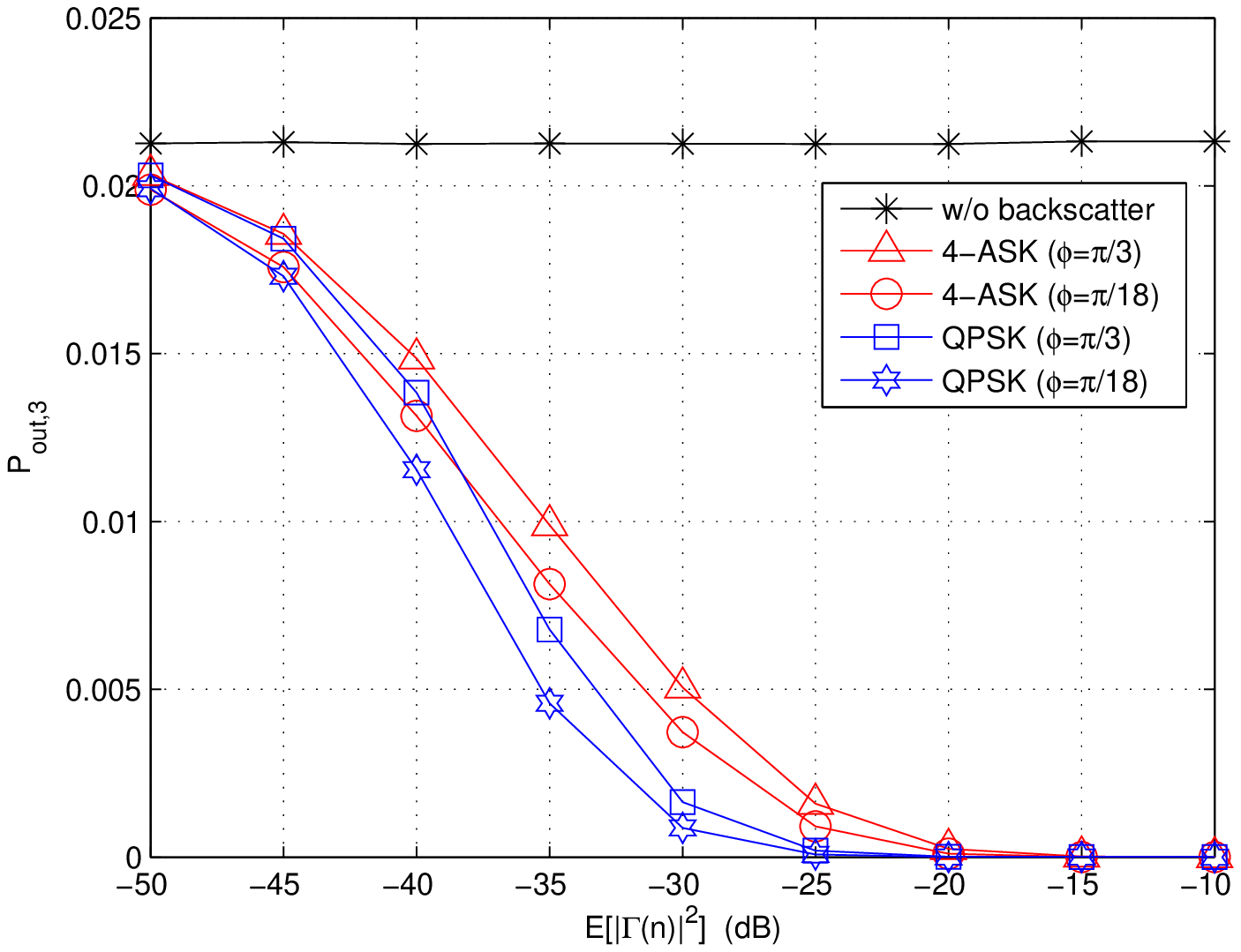}
\caption{Outage probability
of the legacy system versus
$\Es[\left|\Gamma(n)\right|^2]=\alpha^2$ 
for two backscatter signal constellations
and two values of the angle $\phi$.}
\label{fig:fig_5}
\end{minipage}%
\hspace{0.04\textwidth}%
\begin{minipage}[t]{0.45\textwidth}
\includegraphics[width=\linewidth, trim=20 20 40 20]{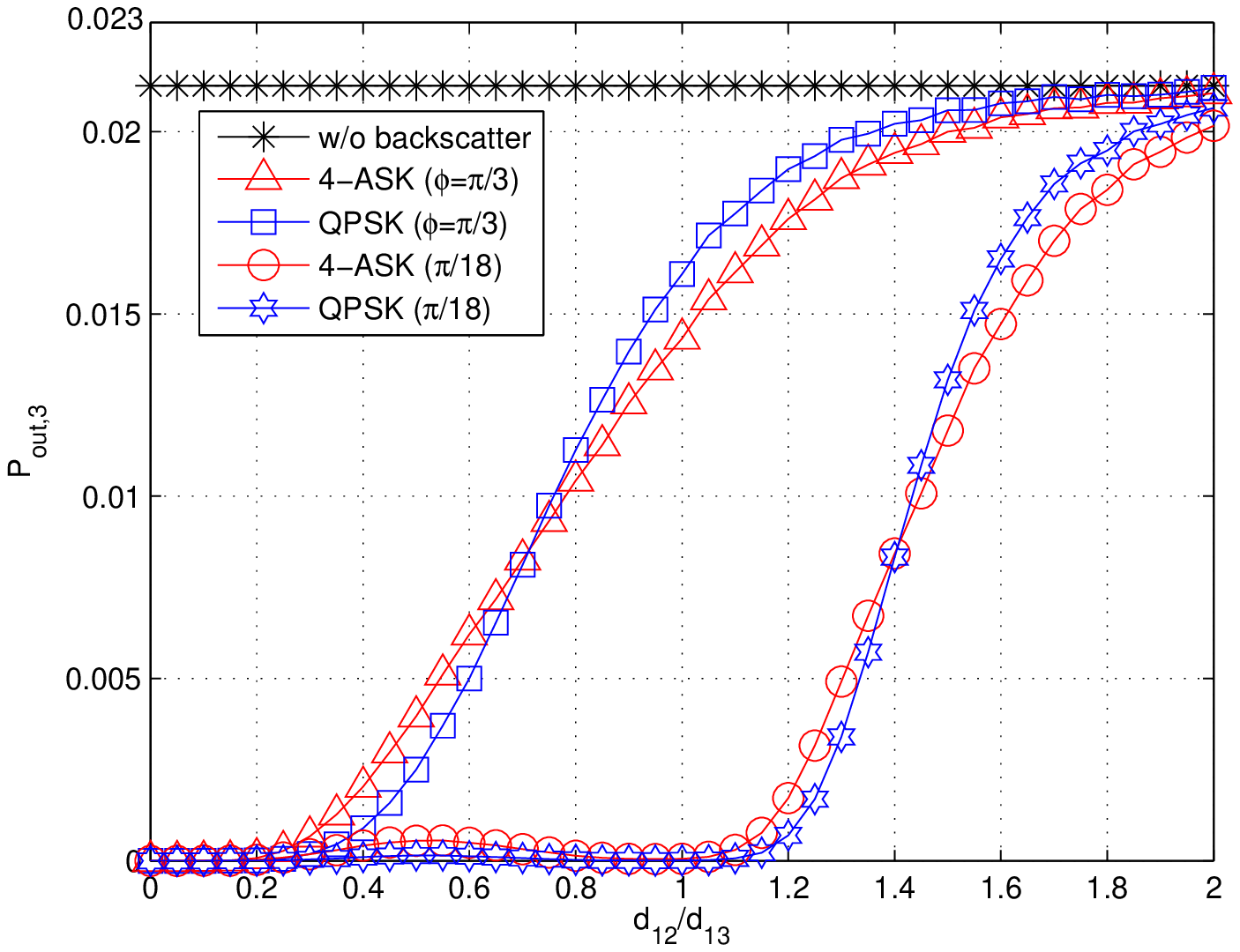}
\caption{Outage probability
of the legacy system versus $d_{12}/d_{13}$ for two backscatter signal constellations  and two values of the angle $\phi$.
}
\label{fig:fig_6}
\end{minipage}
\end{figure*}

\subsection{Performance of the legacy system}

Figs.~\ref{fig:fig_3} and \ref{fig:fig_4} depict the ergodic capacity
$\Cap_3$ of the legacy system given by \eqref{eq:C3-2}, in comparison with
the ergodic capacity \eqref{eq:C3-direct} when the backscatter system
is in sleep mode (referred to as ``w/o backscatter''), with $\SNRL = \sigma_s^2/\sigma_{v_3}^2=20$ dB
and $\phi \in \{\pi/18, \pi/3 \}$.
In Fig.~\ref{fig:fig_3}, the capacity values are reported as
a function of the {\em mean square
power wave reflection
coefficient}
$\Es[\left|\Gamma(n)\right|^2] = \alpha^2 \, \sigma_b^2$, 
with $d_{12}/d_{13}=0.2$, whereas they are plotted
against $d_{12}/d_{13}$ in Fig.~\ref{fig:fig_4},
with $\Es[\left|\Gamma(n)\right|^2]=-20$ dB.

As annunciated in Remark~4, the capacity of the
legacy system cannot degrade in the presence of the
backscatter transmission, in each
operative condition. In particular, the performance gain
$\Delta \Cap_3 =\Cap_3 -\Cap_3 |_{\alpha=0}$ becomes relevant
either when $\Es[\left|\Gamma(n)\right|^2]$  is sufficiently large (see Remark~5)
or the BTx is very close to the LTx.
In particular, it is seen that, for a fixed $Q$, the choice of
the backscatter signal constellation (ASK or PSK)
does not lead to
significantly different values of $\Cap_3$.
Moreover, results of Fig.~\ref{fig:fig_4} confirm the trends
analytically predicted  in Remark~6, by showing that
$\Cap_3$ monotonically decreases as the BTx moves
away from the LTx when $\phi=\pi/3 \in \mathcal{A}$; on the other hand,
when $\phi=\pi/18 \not \in \mathcal{A}$,
the capacity $\Cap_3$ exhibits a local minimum at
$d_{12}/d_{13}=d_{\text{min}}(\pi/18)=0.5252$ and a local maximum
at $d_{12}/d_{13}= d_{\text{max}}(\pi/18)=0.9520$ (i.e., near the LRx).
Similar conclusions can be drawn from the outage probability ${\mathsf{P}}_{\text{out},3}$ given by \eqref{eq:pout3}, as reported in
Figs.~\ref{fig:fig_5} and \ref{fig:fig_6}.

It is important to observe that even small values of $\Delta \Cap_3$
lead to significant increments in terms of data rate for the legacy transmission.
For instance, it can be seen from results of  Fig.~\ref{fig:fig_3}
that, when $\phi=\pi/18$ and the BTx employs a QPSK modulation,
one gets $\Delta \Cap_3= 0.1315$ b/s/Hz
at $\Es[\left|\Gamma(n)\right|^2]=-40$ dB. In this case,
if the LTx is a TV tower broadcasting over a bandwidth of
$6$ MHz \cite{Liu}, then the data-rate gain is
equal to $789$ kbps; on the other hand, if the LTx is
a Wi-Fi access point (AP) operating over a bandwidth of
$20$ MHz \cite{Kello,Bharadia},  the gain is $2.63$ Mbps.

\begin{figure*}[t!]
\centering
\begin{minipage}[t]{0.45\textwidth}
\includegraphics[width=\linewidth, trim=20 20 40 20]{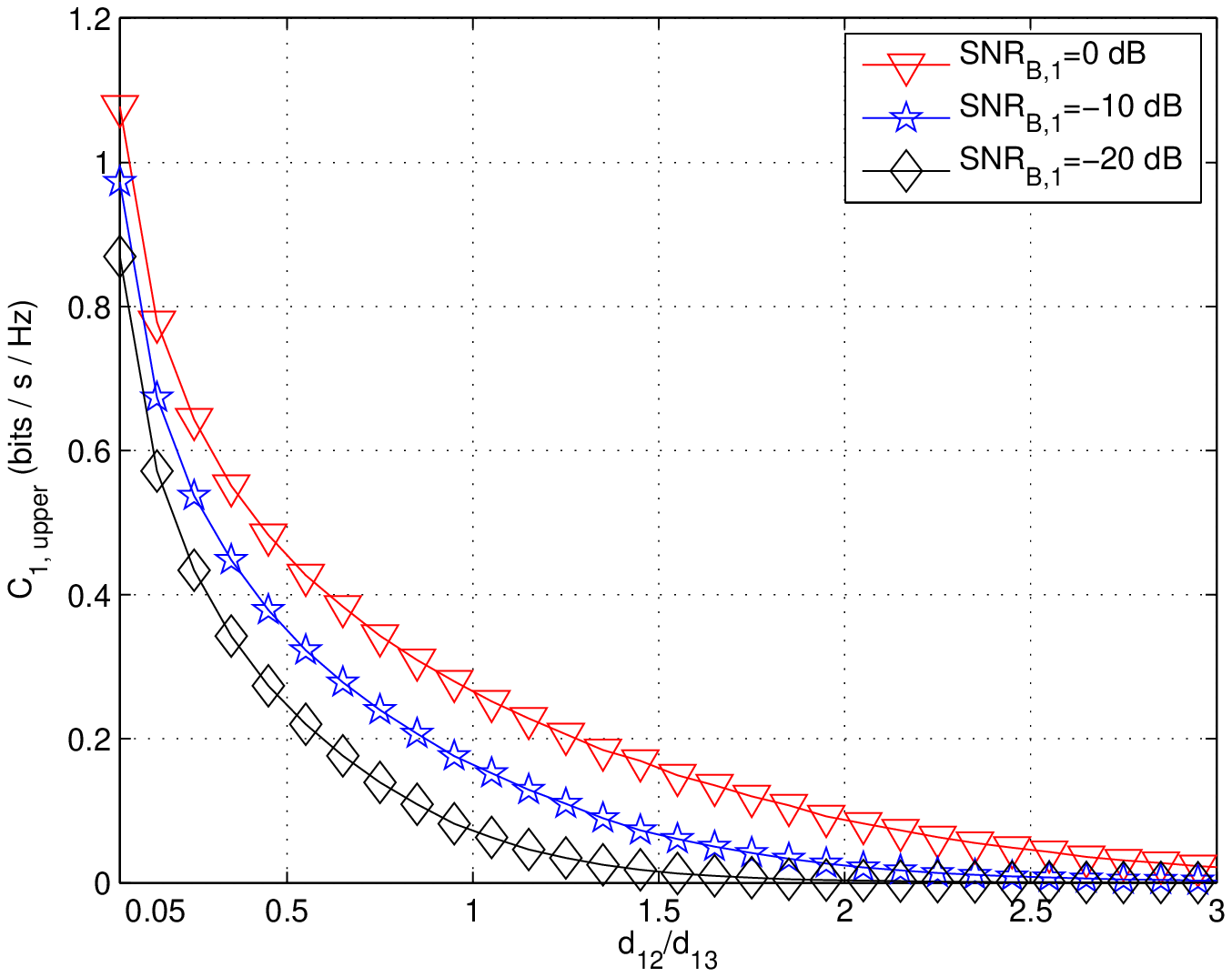}
\caption{Best-case ergodic capacity of the backscatter system versus
$d_{12}/d_{13}$ for different values of $\SNRB$ (LTx and BRx are co-located).}
\label{fig:fig_7}
\end{minipage}%
\hspace{0.04\textwidth}%
\begin{minipage}[t]{0.45\textwidth}
\includegraphics[width=\linewidth, trim=20 20 40 20]{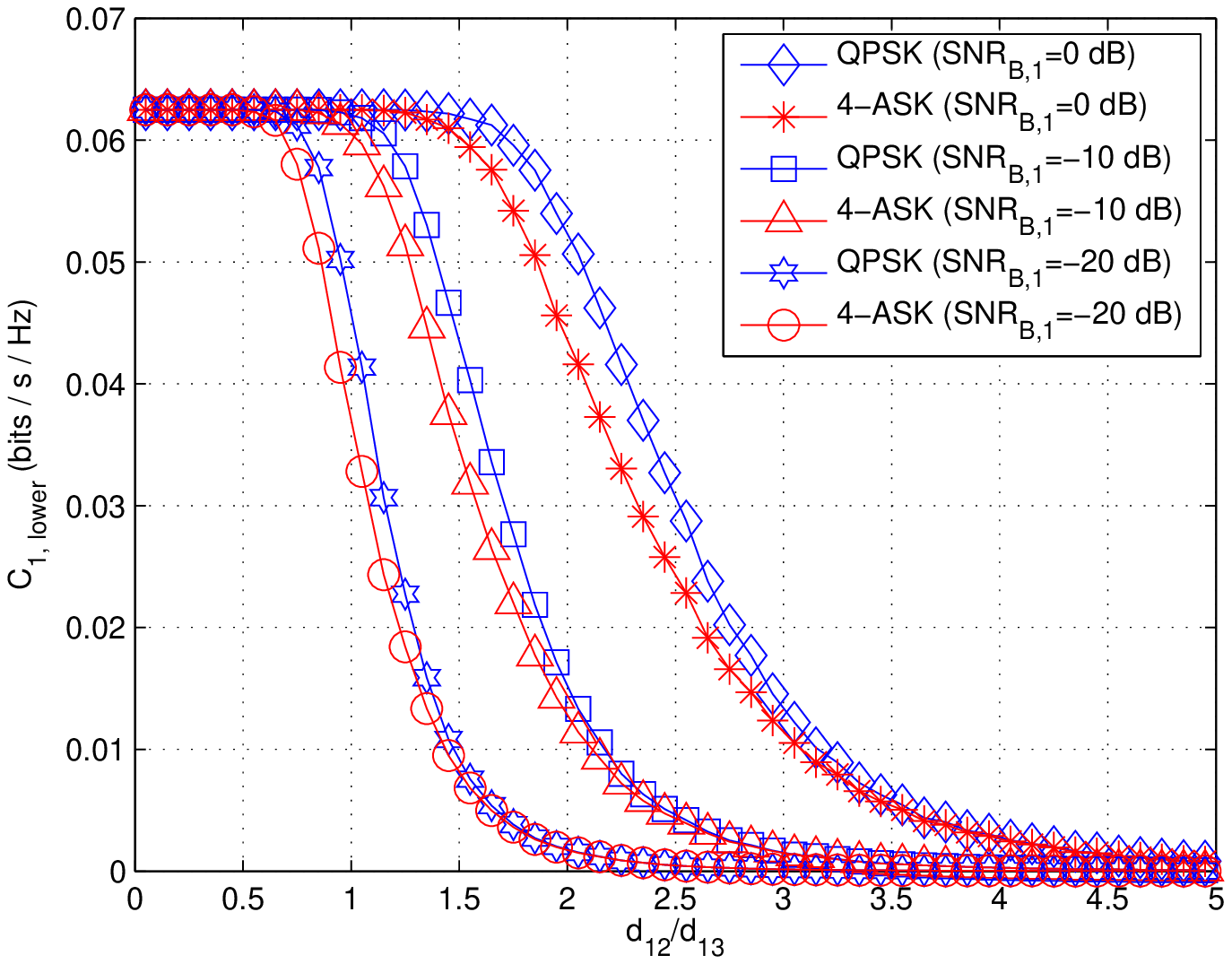}
\caption{Worst-case ergodic capacity of the backscatter system versus
$d_{12}/d_{13}$ for different values of $\SNRB$ (LTx and BRx are co-located).}
\label{fig:fig_8}
\end{minipage}
\end{figure*}

\subsection{Performance of the backscatter system
when the LTx and BRx are co-located}

Herein, we focus on the ergodic capacity $\Cap_1$
of the backscatter system when the intended recipient
BRx of the backscatter transmission is just the energy
source LTx. More precisely, we report in Figs.~\ref{fig:fig_7} and \ref{fig:fig_8}
its upper bound $\Cap_{1,\text{upper}}$ given by
\eqref{eq:C4-1} and lower bound $\Cap_{1,\text{lower}}$ given by
\eqref{eq:lowbouc1} for PSK modulations, respectively, as a function of
$d_{12}/d_{13}$ for different values of $\SNRB$.
We also report in Fig.~\ref{fig:fig_8}
the worst-case ergodic capacity of the backscatter system
for the $4$-ASK case, which is obtained by averaging
\eqref{eq:II11} with respect to $\bpsi$.

As predicted by the performance analysis developed in
Subsection~\ref{sec:coincide}, both the upper and lower
bounds are monotonically decreasing function of
the distance between the LTx and the BTx, for each
value of $\SNRB$.
Moreover, when $d_{12}$ is sufficiently smaller
than $d_{13}$, it results that
$\Cap_{1,\text{lower}} \approx (\log Q)/M=0.0625$, for
each considered value of $\SNRB$.
The slight performance advantage offered
by the QPSK signal constellation over the $4$-ASK one
is due to the fact that the PSK
modulation maximizes the cut-off rate in the
case of equiprobable symbols (see Subsection~\ref{sec:lower}).
Results not reported here show
that the gap between
$\Cap_{1,\text{upper}}$ and $\Cap_{1,\text{lower}}$
is reduced for increasing values of $Q$.

Let us focus on the case when the LTx is a Wi-Fi AP transmitting over
a bandwidth of $20$ MHz,
which might be used to connect the BTx to the Internet \cite{Kello,Bharadia}.
In this scenario, by considering an indoor Wi-Fi network with
$d_{13}=100$ m, we obtain from
Fig.~\ref{fig:fig_8} that the backscatter
communication can achieve at least $1.25$ Mbps
up to a range of $50-70$ m, even for very small values of $\SNRB$.
As a comparison, we underline that
the prototype presented in \cite{Bharadia} is able to 
achieve communication rates up to $1-5$ Mbps at a range of $1-5$ m.
Therefore, compared to \cite{Bharadia}, it is possible in theory to
largely extend the communication range, without 
significantly reducing the data rate.

\begin{figure*}[t!]
\centering
\includegraphics[width=0.45\linewidth, trim=20 40 40 20]{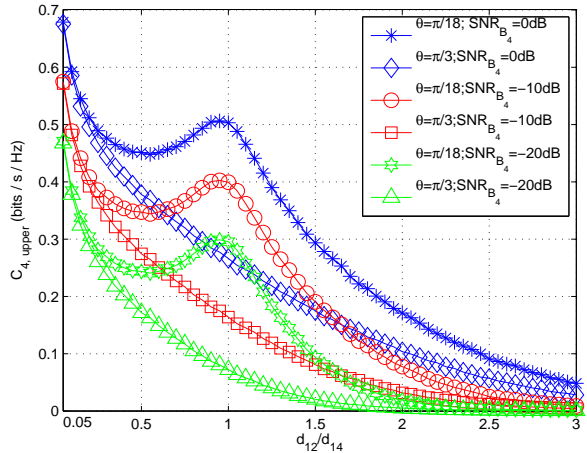}
\caption{Best-case ergodic capacity of the backscatter system versus
$d_{12}/d_{14}$ for different values of $\SNRBB$ (LTx and BRx are 
spatially-separated nodes).}
\label{fig:fig_9}
\end{figure*}

\subsection{Performance of the backscatter system
when the LTx and BRx are spatially-separated nodes}

The last scenario under investigation is when the nodes LTx and BRx
are distinct one from the other,  with
$\theta \in \{\pi/18, \pi/3 \}$ and $d_{14}=1$.
Fig.~\ref{fig:fig_9} depicts the upper bound $\Cap_{4,\text{upper}}$
given by \eqref{eq:C4-11} as a function of
$d_{12}/d_{14}$, for different values of $\SNRBB$.
Results corroborate the discussion
reported in Remark~9, for each value of $\SNRBB$.
In particular, if $\theta=\pi/3 \in \mathcal{A}$,
then  $\Cap_{4,\text{upper}}$ monotonically decreases
as the distance between the BTx and the LTx increases;
when $\theta=\pi/18 \not \in \mathcal{A}$,
the capacity $\Cap_{4,\text{upper}}$ assumes a global maximum
when the BTx tends to be close by the LTx and
a local maximum when the BTx is near the
BRx, i.e., $d_{12}/d_{14}= d_{\text{max}}(\pi/18)=0.9520$,
by taking on a local minimum at
$d_{12}/d_{14}=d_{\text{min}}(\pi/18)=0.5252$.

\begin{figure*}[t!]
\centering
\begin{minipage}[t]{0.45\textwidth}
\includegraphics[width=\linewidth, trim=20 20 40 20]{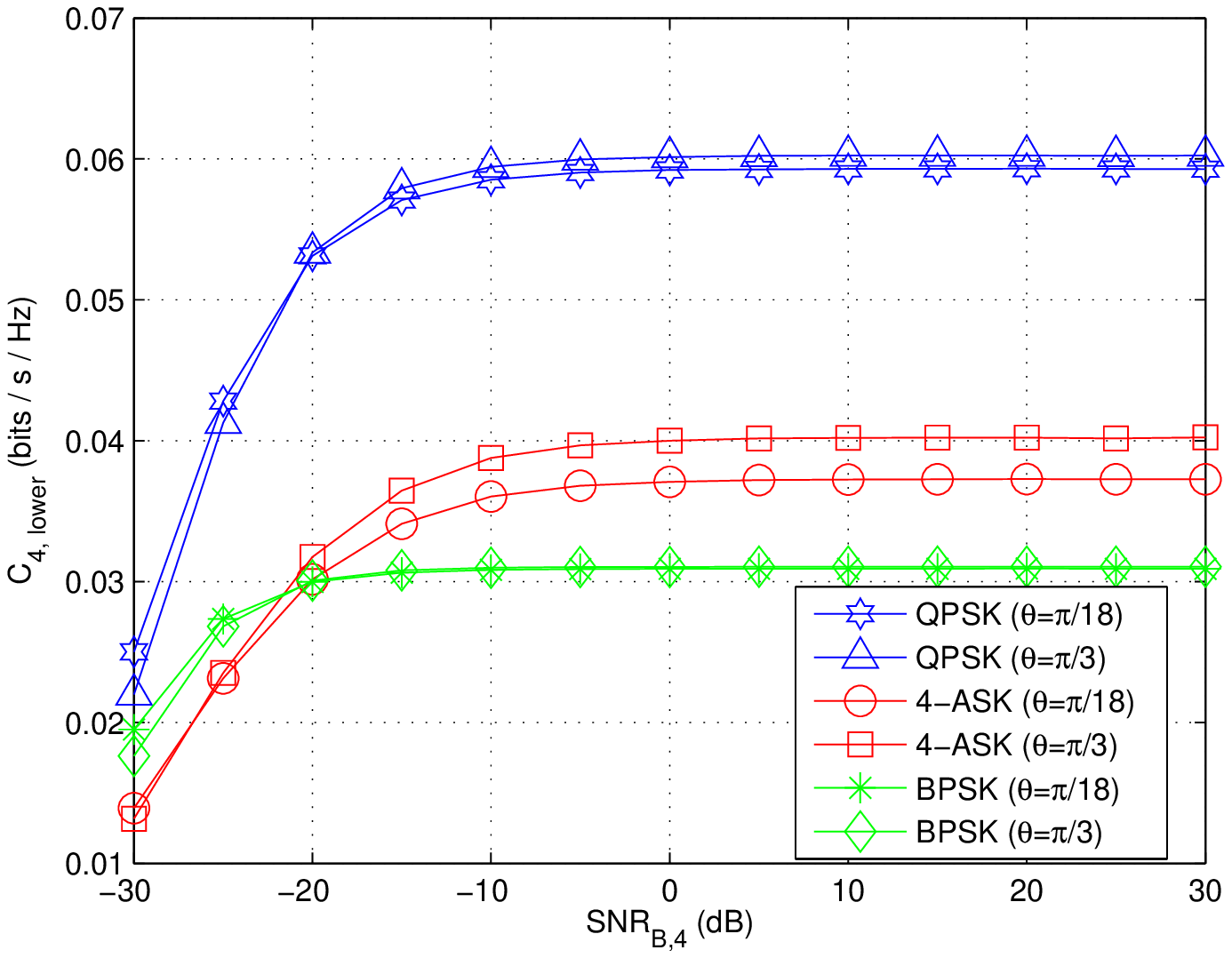}
\caption{Worst-case ergodic capacity of the backscatter system versus
$\SNRBB $ for three backscatter signal constellations
and two values of the angle $\phi$ (LTx and BRx are spatially-separated nodes).}
\label{fig:fig_10}
\end{minipage}%
\hspace{0.04\textwidth}%
\begin{minipage}[t]{0.45\textwidth}
\includegraphics[width=\linewidth, trim=20 20 40 20]{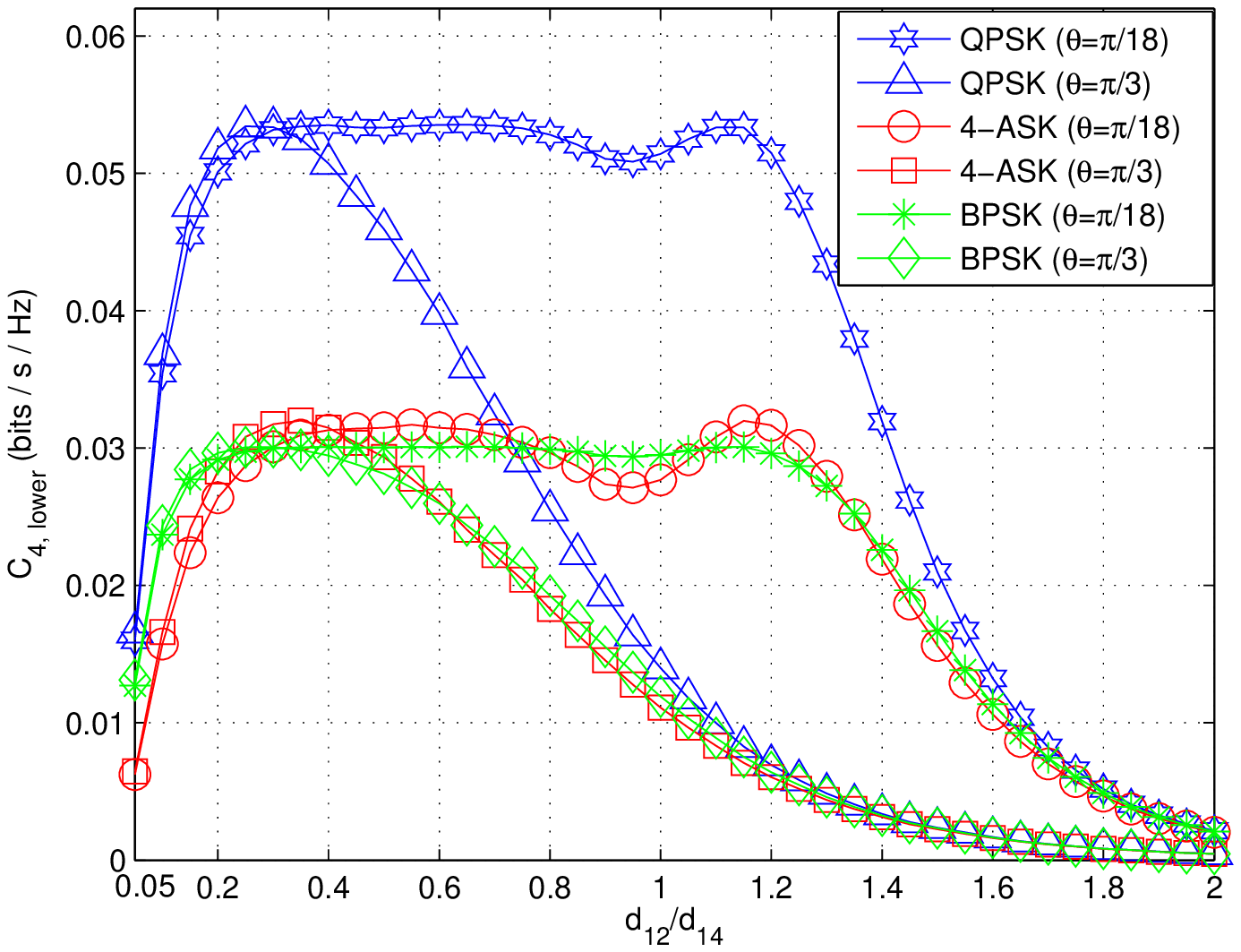}
\caption{Worst-case ergodic capacity of the backscatter system versus $d_{12}/d_{14}$ for two backscatter signal constellations
and two values of the angle $\phi$ (LTx and BRx are spatially-separated nodes).}
\label{fig:fig_11}
\end{minipage}
\end{figure*}

In Fig.~\ref{fig:fig_10}, the capacity $\Cap_{4,\text{lower}}$
given by \eqref{eq:C4-11-low} is reported as
a function of the $\SNRBB$ for different backscatter
signal constellations, with $d_{12}/d_{14}=0.2$,
whereas $\Cap_{4,\text{lower}}$ is
reported in Fig.~\ref{fig:fig_11} as a function of
$d_{12}/d_{14}$, with $\SNRBB=-20$ dB.
It is seen that, also in this case, PSK constellations ensure better
performance in terms of cut-off rate when the symbols are equiprobable.
Another interesting conclusion that can be drawn from
Fig.~\ref{fig:fig_10} is that all curves exhibit a
capacity saturation effect, for vanishingly small noise,
which is due to the interference generated
by the legacy system over the $1 \to 4$ link.
Moreover, independently of the considered backscatter signal constellation,
the capacity $\Cap_{4,\text{lower}}$ is a monomodal function
of $d_{12}/d_{14}$ having a maximum at $d_{12}/d_{14} \approx 0.3$
when $\theta=\pi/3 \in \mathcal{A}$, whereas, for $\theta=\pi/18 \not \in \mathcal{A}$,  it is multimodal by exhibiting
slight fluctuations over a large interval of distances
ranging from $d_{12}/d_{14} \approx 0.2$ to $d_{12}/d_{14} \approx 1.2$.

Let us consider the practical scenario when
the LTx is a TV tower broadcasting over a bandwidth of
$6$ MHz \cite{Liu}, with $d_{14}=4$ Km. In this case,
according to the results of Fig.~\ref{fig:fig_10},
by employing a QPSK backscatter signal constellation,
the worst-case achievable data rate is equal to
$360$ kbps over a distance of $800$ m
at  $\SNRBB=-10$ dB.
As a comparison, we underline that
the prototype presented in \cite{Liu} is able to 
achieve information rates of $1$ kbps over a 
distance of $5-8$ m.
Therefore, compared to \cite{Liu}, it is possible in theory to
significantly extend both the communication range and the data rate.

\section{Conclusions}
\label{sec:concl}

We developed a general framework for evaluating the
ultimate achievable rates of a point-to-point backscatter communication network,
by considering the influence of the backscatter
transmission on the performance of the legacy system,
from which energy is opportunistically harvested.
Our theoretical results show that,
in principle, ambient backscatter allows a passive device to achieve significant
communication rates over short distances. As a by-product, the
backscatter transmission can even ensure a performance improvement of
the legacy system, provided that the latter one is designed to
exploit the additional diversity arising from the backscatter process.

In view of the prototypes and experiments presented
in \cite{Liu,Kello,Bharadia}, we highlight that there
is plenty of scope for performance improvement, which
mandates the use of advanced signal processing techniques,
especially at the intended recipient of the backscatter information. Moreover,
results of our performance analysis pave the way towards various
system-level optimizations. Among the others, an interesting issue
is to analytically determine what is the optimal choice of $\Es[\left|\Gamma(n)\right|^2]$ that
ensures the best tradeoff between performance of
legacy/backscatter systems and energy harvesting at the
passive backscatter transmitter.

% %^^^^^^^^^^^^^^^^^Bibliography^^^^^^^^^^^^^^^^^^^^^^^^^^^^^^^^^^


\begin{thebibliography}{99}

\setlength{\baselineskip}{5.0mm}
\itemsep=-1mm

\bibitem{Boy2014}
C.~Boyer and S.~Roy,
``Backscatter communication and RFID: Coding, Energy, and MIMO analysis'',
\IeeeTCOMM, pp.\ 770--785, Mar.\ 2014.

\bibitem{Dar2015}
D.~Darsena, G.~Gelli, and F.~Verde,
``Exploiting noncircularity in backscattering communications'',
in {\em Proc.\ of the Twelfth International Symposium on Wireless Communication Systems (ISWCS)\/}, Brussels, Belgium,
Aug.\ 2015, pp.\ 1-5.

\bibitem{Liu}
V.~Liu, A.~Parks, V.~Talla, S.~Gollakota, D.~Wetherall, and J.R.~Smith,
``Ambient backscatter: wireless communication out of thin air'',
 in {\em Proc. of ACM SIGCOMM'13},
 Hong Kong, China, Aug.\ 2013, pp.\ 39--50.

\bibitem{Kello}
B.~Kellogg, A.~Parks, S.~Gollakota, J.R.~Smith, and D.~Wetherall,
``Wi-Fi backscatter: Internet connectivity for RF-powered devices'',
 in {\em Proc. of ACM SIGCOMM'14},
 Chicago, Illinois, USA, Aug.\ 2014, pp.\ 607--618.

\bibitem{Bharadia}
D.~Bharadia, K.~Joshi, M.~Kotaru, S.~Katti,
``BackFi: High throughput WiFi backscatter'',
 in {\em Proc. of ACM SIGCOMM'15},
 London, United Kingdom, Aug.\ 2015, pp.\ 283--296.

\bibitem{Ma2015}
Z.~Ma, T.~Zeng, G.~Wang, and F.~Gao,
``Signal detection for ambient backscatter system with multiple receiving antennas'',
in {\em Proc.\ of IEEE 14th Canadian Workshop on
Information Theory (CWIT)\/}, St.\ John's, NL, Canada,
July 2015, pp.\ 50-53.

\bibitem{Lu2015}
K.~Lu, G.~Wang, F.~Qu, and Z.~Zhong,
``Signal detection and BER analysis for RF-powered devices utilizing ambient backscatter'',
in {\em Proc.\ of  International Conference on Wireless Communications \& Signal Processing (WCSP)\/}, Nanjing, China,  Oct.\ 2015, pp.\ 1-5.

\bibitem{Wang2015}
G.~Wang, F.~Gao, Z.~Dou, and C.~Tellambura,
``Uplink Detection and BER Analysis for Ambient Backscatter Communication Systems'',
in {\em Proc.\ of IEEE Global Communications Conference (Globecom)\/},
San Diego, CA, USA, Dec.\ 2015, pp.\ 1-6.

\bibitem{Sha}
H.~Shariatmadari {\em et al.}, ``Machine-type communications: Current status and future perspectives toward 5G systems'', \IeeeCOMMMAG,
pp.\ 10--17, Sep.\ 2015.

\bibitem{Ash}
A.~Sabharwal {et al.}, ``In-band full-duplex wireless: Challenges and
opportunities,'' \IeeeJSAC,
vol.\ 32, pp.\ 1637--1652, Sep.\ 2014.

\bibitem{Horn}
R.~A.~Horn and C.~R.~Johnson,
{\em Matrix Analysis\/},
Cambridge: Cambridge University Press, 1990.

\bibitem{Wang}
Z.~Wang and G.B.~Giannakis, ``Wireless multicarrier communications --
where Fourier meets Shannon,'' \IeeeSPMAG,
pp.\ 29--48, May 2000.

\bibitem{Stock}
H.~Stockman,``Communication by means of reflected power'',
{\em Proc.\ IRE}, pp.\ 1196--1204, Oct.\ 1948.

\bibitem{Tho}
S.J.~Thomas, E.~Wheeler, J.~Teizer, and M.S.~Reynolds,
``Quadrature amplitude modulated backscatter in passive and
semipassive UHF RFID systems'',
\IeeeTMTT, pp.\ 1175--1182, Apr.\ 2012.

\bibitem{King}
D.D.~King,``The measurement and interpretation of antenna scattering'',
{\em Proc.\ IRE}, pp.\ 770--777, July 1949.

\bibitem{Kuro}
K.~Kurokawa,
``Power waves and the scattering matrix'',
\IeeeTMTT, pp.\ 194--202, Mar.\ 1965.

\bibitem{Arn}
D.~Arnitz, U.~Muehlmann, and K.~Witrisal,
``Tag-based sensing and positioning in passive UHF RFID: Tag reflection'',
 in {\em Proc. of 3rd Int.\ EURASIP Workshop RFID
Technol.\/}, Cartagena, Spain, Sep.\ 2010, pp.\ 51--56.

\bibitem{Han}
R.C.~Hansen,``Relationships between antennas as scatters and radiators'',
{\em Proc.\ IEEE}, pp.\ 659--662, May 1969.

\bibitem{Mengali_book}
U.~Mengali and A.N.~D'Andrea,
{\em Synchronization Techniques for Digital
Receivers}, New York: Plenum, 1997.

\bibitem{Morelli}
M.~Morelli, C.-C.J.~Kuo, and M.-O.~Pun, ``Synchronization techniques
for orthogonal frequency division multiple access (OFDMA): a tutorial
review," {\em Proc.\ IEEE}, vol.\ 95, pp.\ 1394--1427, July 2007.

\bibitem{Picinbono}
B.~Picinbono, ``On circularity,'' \IeeeTSP, vol.\ 42,
pp.\ 3473--3482, Dec.\ 1994.

\bibitem{Telatar}
I.E.~Telatar, ``Capacity of multi-antenna Gaussian channels,''
{\em Eur.\ Trans.\ Telecommun.\/}, vol.\ 10,
pp.\ 585--595, Nov./Dec.\ 1999.

\bibitem{Gallager_book}
R.G.~Gallager, {\em Information Theory and Reliable Communication},
New York: Wiley, 1968.

\bibitem{CoverThomas_book}
T.M.~Cover and J.A.~Thomas, {\em Elements of Information Theory},
New York: Wiley, 1991.

\bibitem{Biglieri}
E.~Biglieri, J.G.~Proakis, and S.~Shamai, ``Fading Channels: Information-Theoretic and Communications Aspects,'' \IeeeTIT,
vol.\ 44, pp.\ 2916--2692, Oct.\ 1998.

\bibitem{Ozarow}
L.~Ozarow, S.~Shamai, and A.~Wyner,
``Information theoretic considerations for cellular mobile radio,''
\IeeeTVT, vol.\ 43, pp.\ 359 -- 378, May 1994.

\bibitem{Mengali}
M.~Morelli and U.~Mengali, ``A comparison of pilot-aided channel
estimation methods for OFDM systems,'' \IeeeTSP, pp.\ 3065--3073, Dec.\ 2001.

\bibitem{Smith}
J G.~Smith, ``The information capacity of amplitude and 
variance-constrained scalar Gaussian channels," {\em Inf.\ Contr.\/}, pp.\ 203--219, 1971.

\bibitem{Nee}
F.D.~Neeser and J.L.~Massey, ``Proper complex random processes
with applications to information theory,''
\IeeeTIT, pp.\ 1293--1302, July 1993.

\bibitem{Wu}
Y.~Wu and S.~Verd\'{u},
``The impact of constellation cardinality on Gaussian channel capacity,"
in {\em Proc.\ of the 48th Annual Allerton Conference on
Communication, Control, and Computing (Allerton)\/}, Allerton, IL, USA,
Sep.-Oct. \ 2010, pp.\ 620--628.

\bibitem{Papoulis}
A.~Papoulis.
{\em Probability, Random variables, and Stochastics Processes (3rd ed.).}
McGraw-Hill, Singapore, 1991.

\bibitem{Wilson}
S.G.~Wilson.
{\em Digital Modulation and Coding.}
Englewood Cliffs, NJ: Prentice Hall, 1996.

\bibitem{Marzetta1999}
T.L.~Marzetta and B.M.~Hochwald, ``Capacity of a mobile multiple-antenna
communication link in Rayleigh flat fading,'' \IeeeTIT,
vol.\ 45, pp.\ 139--157, Jan.\ 1999.

\bibitem{Hock2000}
B.M.~Hochwald and T.L.~Marzetta, ``Unitary space-time modulation for
multiple-antenna communications in Rayleigh flat fading,'' \IeeeTIT,
vol.\ 46, pp.\ 543--564, Mar.\ 2000.

\bibitem{Zheng2002}
L.~Zheng and D.N.C.~Tse, ``Communication on the Grassmann Manifold: A geometric approach to the noncoherent multiple-antenna channel,'' \IeeeTIT,
vol.\ 48, pp.\ 359--383, Feb.\ 2002.

\end{thebibliography}
\end{document}